\documentclass[12pt]{article}
\usepackage[OT1]{fontenc}
\usepackage[utf8]{inputenc}
\usepackage[margin=0.9in]{geometry}
\usepackage{graphicx}
\usepackage{algorithm}
\usepackage{algorithmic}
\usepackage{amsmath}
\usepackage{amsfonts}
\usepackage{comment} 
\usepackage{appendix}
\usepackage{subfigure}
\usepackage[usenames,dvipsnames]{xcolor}
\usepackage{lipsum}  
\usepackage{tikz}
\usepackage{amsthm}
\usepackage[affil-it]{authblk}

\newcommand\nst[1]{{\color{black}#1}} 
\newcommand\reva[1]{{\color{black}#1}}
\newcommand\revb[1]{{\color{black}#1}}
\newcommand\revc[1]{{\color{black}#1}}

\newcommand{\beginsupplement}{%
        \setcounter{table}{0}
        \renewcommand{\thetable}{S\arabic{table}}%
        \setcounter{figure}{0}
        \renewcommand{\thefigure}{S\arabic{figure}}%
     }

\title{Meta-Learning Biologically Plausible Plasticity Rules with Random Feedback Pathways}
\author[*,1]{Navid Shervani-Tabar}
\author[1]{Robert Rosenbaum}
\affil[1]{Department of Applied and Computational Mathematics and Statistics, University of Notre Dame, Notre Dame, IN 46556, USA}
\affil[*]{Corresponding author: nshervan@nd.edu}
\date{\today}

\begin{document}
\maketitle    

\begin{abstract}    
    Backpropagation is widely used to train artificial neural networks, but its  relationship to synaptic plasticity in the brain is unknown. Some biological models of backpropagation rely on feedback projections that are symmetric with feedforward connections, but experiments do not corroborate the existence of such symmetric backward connectivity. Random feedback alignment offers an alternative model in which errors are propagated backward through fixed, random backward connections. This approach successfully trains shallow models, but learns slowly and does not perform well with deeper models or online learning. In this study, we develop a \reva{meta-learning} approach to discover interpretable, biologically plausible plasticity rules that improve online learning performance with fixed random feedback connections. The resulting plasticity rules show improved online training of deep models in the low data regime. Our results highlight the potential of \reva{meta-learning} to discover effective, interpretable learning rules satisfying biological constraints.
\end{abstract}

\section{Introduction}

Error-driven learning in multilayer neural networks was revolutionized by the error backpropagation algorithm~\cite{rumelhart1986learning}, or backprop for short. In backprop, gradients or ``errors'' are propagated backward through auxiliary feedback pathways to compute parameter updates. 

While practical, backprop has strong structural constraints that make it biologically implausible~\cite{whittington2019theories,lillicrap2020backpropagation}. A major limitation, known as the weight transport problem~\cite{grossberg1987competitive} states that transmitting gradients to upstream layers requires feedback connections that are symmetric with feedforward connections. Such symmetric connectivity is not known to exist in the brain. 
In an attempt to depart from the symmetry assumption, Lillicrap~\textit{et al.}~\cite{lillicrap2016random} show that even random backward connections can transmit effective teaching signals to train the upstream layers. In this scenario, while the backward connections are fixed, forward weights evolve to align the teaching signals with those prescribed by the backprop algorithm. However, leaving out the symmetry constraint comes with caveats. Random feedback alignment struggles with deeper networks, limited training data sizes, convolutional layers, and online data streams~\cite{amit2019deep, bartunov2018assessing}.
 

To improve random feedback alignment, Nøkland~\cite{nokland2016direct} proposed to rewire the feedback connections and feed the teaching signals directly from the output layer to the upstream layers. While this improves the transmission of errors, it still does not perform as robustly as the symmetric case in the low data regime. Parallel to this, Liao~\textit{et al.}~\cite{liao2016important} suggested dismissing symmetry in magnitude, but assigning symmetric signs to the feedback connections. Nonetheless, they found that decreasing the batch size of the training data may deteriorate performance when discarding symmetry. In addition, they found batch normalization~\cite{ioffe2015batch} critical for training with asymmetric connections. These findings render the methods inadequate for training with an online stream of data, where the batch size is one, and ultimately undercuts their biological plausibility.

An alternative strategy is to implement a secondary update rule to modify the backward connections along with the forward weights. To that end, Akrout~\textit{et al.}~\cite{akrout2019deep} proposed to use a Hebbian plasticity rule~\cite{hebb2005organization} to adjust the feedback matrices parallel to the approximate gradient-based update of the forward path. The former pushes the backward connections toward the transpose of the forward weights. However, Kunin~\textit{et al.}~\cite{kunin2020two} show that this approach is highly sensitive to hyperparameter tuning. Instead, they redefine the optimization objective as a loss function based on the forward path in combination with layer-wise regularization terms for backward weights to update forward and backward pathways concurrently. They propose a few regularization terms and show that combining these units can achieve more stable plasticity rules.

Meta-learning is a broad learning framework consisting of a learning process that envelopes another optimization loop and learns some aspect of the inner learning procedure, effectively ``learning to learn.'' Although this concept has been around for decades~\cite{schmidhuber1992learning}, Finn~\textit{et al.}~\cite{finn2017model} popularized meta-learning for  few-shot learning applications. This approach employs meta-learning to optimize an internal representation of the network, which is subsequently used as an initial weight to expedite learning on a downstream task. Further, Javed and White~\cite{javed2019meta} extended this approach to continual learning by modifying the objective function of the outer optimization loop. Still, they used the modified approach to learn a partial initialization of the network's forward weights. Although effective in learning representations for the few-shot learning, they effectively work by pre-training a model rather than by learning to learn. More precisely, their effectiveness is largely derived from their ability to meta-learn a weight initialization, rather than meta-learning a learning rule itself.

The meta-learning framework has provided a new direction for building biologically plausible computational neural models. For example, Lindsey~\textit{et al.}~\cite{lindsey2020learning} learn the direct feedback pathways that modulate activations and use a supervised adaptation of Oja's rule to update forward connections. It is supervised because it benefits from modulated activations, which do not guarantee the established properties of conventional Oja's rule~\cite{oja1982simplified}. Nevertheless, they also meta-learn an initial value for the forward connections, which makes their approach dependent on the learned weight initialization, not on the learned learning rule alone. Miconi~\textit{et al.}~\cite{miconi2018differentiable,miconi2018backpropamine} showed that \reva{meta-learning} can train a variety of network architectures on various tasks. Like Lindsey~\textit{et al.}~\cite{lindsey2020learning}, their approach meta-trains a separate plasticity rule for each weight. While this approach can be effective, the resulting plasticity rules are difficult to interpret. In addition, meta-learning weight initialization in these works makes it unclear to what degree the results are affected by the proposed plasticity rule as opposed to the weight initialization.

A growing body of work aims to only meta-learn a plasticity rule without inferring any component of the inner model, such as initial weights. Early work includes Bengio~\textit{et al.}~\cite{bengio1995search}, who meta-learned a parametric learning rule to train a 2D classifier and boolean function. In each meta-iteration, they used the plasticity rule to train multiple networks on separate tasks and obtained the meta-loss function by summing over the loss of all these networks. More recent work includes \reva{Andrychowicz~\textit{et al.}~\cite{andrychowicz2016learning}, who parametrize the learning rule with a Recurrent Neural Network (RNN) and meta-learn weights of the RNN model. Using an RNN allows for training a dynamic update rule. In the context of biological plausibility,} Confavreux~\textit{et al.}~\cite{confavreux2020meta} used \reva{meta-learning} to determine plasticity rules that train shallow linear networks. Rather than discovering new rules, they recover well-known plasticity rules using objective functions based on their known behavior.


\reva{The scope of the meta-learning framework is beyond learning the forward pathway's plasticity rule. Meta-learning has given rise to unorthodox training models beyond the classic backward transmission of errors. For example, Metz~\textit{et al.}~\cite{metz2018meta} used a meta-learning framework to learn a plasticity rule for unsupervised learning. They proposed to infer the teaching signals by meta-learning a network that projects forward activation units and the downstream feedback signal into backward hidden states. These hidden states are subsequently used to update the forward and backward weights via each pathway's meta-learned plasticity rule. Another related work on semi-supervised learning~\cite{gu2019meta} uses learnable auxiliary feedback and lateral connections to facilitate error propagation during training and meta-learns the plasticity rules to update these connections. Finally, Sandler~\textit{et al.}~\cite{sandler2021meta} reformulate the interactions between the forward and backward activations by defining parameterized update rules for both feedforward and feedback connections. Then, they yield new plasticity rules by meta-learning these hyperparameters.}

Here, we improve upon previous work by discovering a plasticity rule that enhances the flow of information in the backward pathway while learning more distinctive embeddings in the forward network. We use \reva{meta-learning} to learn a parameterized plasticity rule based on a combination of \revb{candidate} rules. Key features that characterize our approach include:
\begin{enumerate}
    \item Our approach solely meta-learns a plasticity rule and does not learn a weight initialization. As a result, our approach learns a learning rule that can be applied to train ``naive,'' randomly initialized networks from scratch.
    \item We use ``meta-parameter sharing'' in the sense that all weights share a common plasticity rule, instead of learning a separate plasticity rule for each weight. This approach allows us to interpret and understand the meta-learned plasticity rules.
    \item We impose an $L1$ penalty on the plasticity coefficients in our meta-loss function. This encourages our algorithm to learn a plasticity rule with fewer terms, further simplifying the analysis and interpretability of the resulting rule.
    \item Our inner learning loop uses online learning (batch size $1$) and limited training data ($250$ data points). Coupled with the random weight initialization in our inner learning loop, this  forces the plasticity rules to learn in a more biologically relevant and challenging setting, \revb{with which random feedback alignment is known to struggle~\cite{liao2016important}.}
\end{enumerate}



\revb{
Previous studies have employed different combinations of elements, such as meta-parameter sharing (as used in~\cite{confavreux2020meta}) and online learning (as used in~\cite{lindsey2020learning}). In contrast to previous studies, we integrate all these features to address the weight alignment problem. Our analysis of the meta-learned plasticity rules demonstrates how they overcome the weight alignment challenge.}  
Our approach further advances the use of meta-plasticity to understand how effective learning can emerge in biological neural circuits. 




\section{Results}
\subsection{Feedback Alignment does not learn effectively in deep networks\label{sec:err}}
Consider a fully-connected deep neural network $f_{\boldsymbol{W}}$ parameterized by weights $\boldsymbol{W}$, representing a non-linear mapping $f_{\boldsymbol{W}}:\boldsymbol{x}\mapsto\boldsymbol{y}_L$ from the network's input $\boldsymbol{y}_0=\boldsymbol{x}$ to the output $\boldsymbol{y}_L$, with $L$ denoting the depth of the network. Each network layer is defined by

\begin{align}
    \boldsymbol{z}_{\ell}&=\boldsymbol{W}_{\ell-1,\ell}\boldsymbol{y}_{\ell-1},
    \label{eq:z}\\
    \boldsymbol{y}_{\ell}&=\sigma(\boldsymbol{z}_{\ell}),
    \label{eq:y}
\end{align}
where $\boldsymbol{y}_{\ell}$ is the activation for layer $\ell$ and $\sigma$ stands for the non-linear activation function.

Given a dataset $\mathcal{D}_{train}=(\boldsymbol{X}_{train}, \boldsymbol{Y}_{train})$, the model is trained in an attempt to find the set of weight parameters $\boldsymbol{W}=\{\boldsymbol{W}_{\ell-1,\ell}|0<\ell\leq L\}$, that minimize a loss function $\mathcal{L}(\boldsymbol{y}_{L}, \boldsymbol{Y}_{train})$. 
\revb{Each weight matrix $\boldsymbol{W}_{\ell-1,\ell}$ is modulated by a teaching signal $\boldsymbol{e}_{\ell}$ derived from $\mathcal{L}$. A commonly used method to compute $\boldsymbol{e}_{\ell}$ is to analytically calculate the modulatory signal $\boldsymbol{e}_L$ in the output layer and then use a backward auxiliary network to transmit it to the upstream layers. This backward projection follows the relation
\begin{equation}
    \boldsymbol{e}_{\ell}= \boldsymbol{B}_{\ell+1,\ell} \boldsymbol{e}_{\ell+1}\odot\sigma'(\boldsymbol{z}_{\ell}),
    \label{eq:e_l_backprop}
\end{equation}
where $\odot$ denotes element-wise multiplication and $\boldsymbol{B}=\{\boldsymbol{B}_{\ell+1,\ell}|0<\ell<L\}$ are the set of feedback connections.} 

\revb{In a gradient-based optimization algorithm, $\boldsymbol{e}_L$ is defined as the derivative of the loss function $\mathcal L$ with respect to $\boldsymbol{z}_L$. This teaching signal is propagated backward up to the initial layer to modulate the weight parameters. A widely used scheme, backprop, uses feedback weights $\boldsymbol{B}^{BP}_{\ell+1, \ell}$ that are the transposes of the forward path's weights, to transport these modulating signals using Eq.~\ref{eq:e_l_backprop}. Subsequently, the forward} weight parameters are updated by
\begin{equation}
    \Delta \boldsymbol{W}_{\ell-1, \ell}=-\theta \boldsymbol{e}_{\ell}\boldsymbol{y}_{\ell-1}^T,
    \label{eq:F_grad}
\end{equation}
which represents a shared plasticity rule for all forward connections $\boldsymbol{W}_{\ell-1, \ell}$ and $\theta$ is the associated learning rate. 

To alleviate the biologically undesirable characteristics of the backprop algorithm, Ref.~\cite{lillicrap2016random} proposed the ``Random Feedback Alignment'' approach, which departs from the assumption of symmetric feedback connections \revb{and} instead uses fixed random backward connections \revb{$\boldsymbol{B}^{FA}$} that are not bound to the forward weights. 
To distinguish between the two learning algorithms, we hereafter use the phrase ``feedback alignment'' to refer to the learning rule in Eq.~\ref{eq:F_grad} with fixed random \revb{$\boldsymbol{B}^{FA}_{\ell+1,\ell}$} and we use ``backprop'' to refer to Eq.~\ref{eq:F_grad} with $\revb{\boldsymbol{B}^{BP}_{\ell+1, \ell}}=\boldsymbol{W}_{\ell, \ell+1}^T$.

\begin{figure}[H]
    \centering
    \includegraphics[width=.97\textwidth]{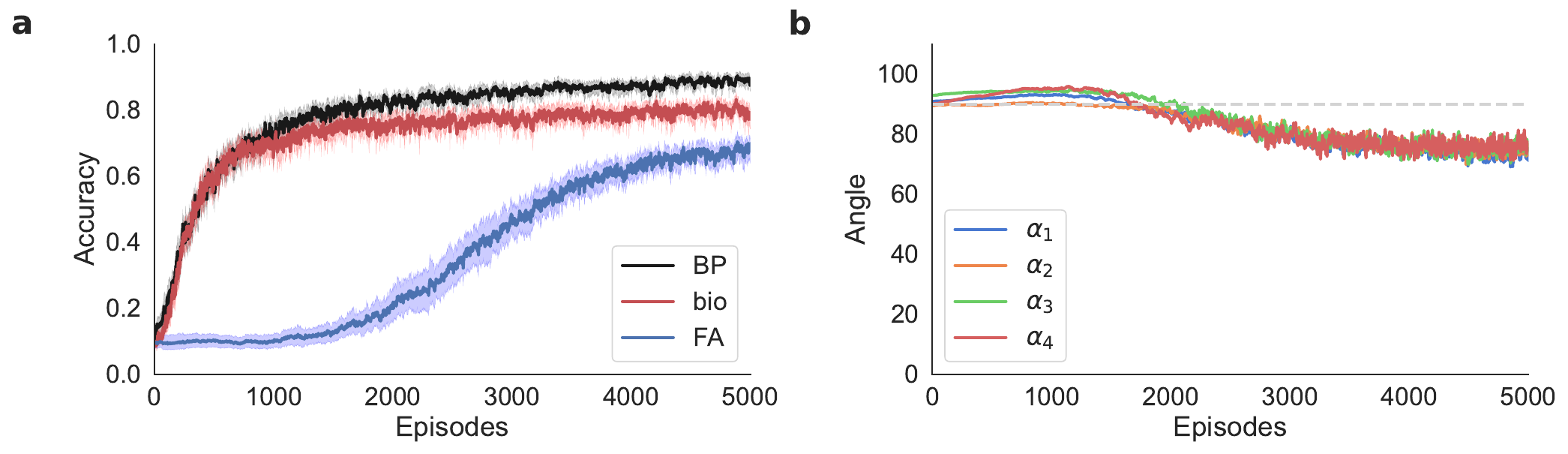}
    \caption{\textbf{Feedback alignment learns poorly in deep models:} \footnotesize \textit{Performance of benchmark learning schemes while training a $5-$layer fully-connected classifier network on MNIST digits~\cite{lecun1998gradient} with online learning. (a) Accuracy versus the number of training data for Feedback Alignment (FA)~\cite{lillicrap2016random} and backprop (BP)~\cite{rumelhart1986learning} methods, compared to the discovered biologically \revb{plausible} plasticity rule (bio) in Sec.~\ref{subsec:F_bio}. (b) The angle $\alpha_{\ell}$ between the teaching signal $\boldsymbol{e}^{FA}_{\ell}$ transmitted by the Feedback Alignment method and the corresponding backpropagated signal $\boldsymbol{e}^{BP}_{\ell}$. \revc{In all figures, each plot illustrates the mean over multiple trials. The shaded area represents the $98\%$ confidence interval (see Methods).}\label{fig:benchmarks}}}
\end{figure}

For feedback alignment, the teaching signal $\boldsymbol{e}^{FA}_{\ell}$ is not an exact gradient, but an approximating pseudo-gradient term. The resulting learning algorithm performs well on simple tasks and shallower networks. However, feedback alignment fails to reach good accuracy in deeper networks and is not as robust in the small data regime. 
In our empirical test with an online stream of data, feedback alignment \reva{only} begins to effectively learn after about 2000 iterations, while backprop learns much more quickly (Fig.~\ref{fig:benchmarks}a). 
\revb{An alternative approach to using feedback connections that link consecutive layers is to create direct backward pathways~\cite{nokland2016direct}. This change allows errors to be transmitted directly from the output layer to the upstream layers. This modification leads to improved performance compared to the feedback alignment method, speeding up the learning process and improving accuracy. However, it still falls short of the performance level of backpropagation (see Supplementary Fig.~\ref{fig:DFA}).}
In addition, Fig.~\ref{fig:benchmarks}b shows that the teaching signals transmitted through fixed feedback connections $\boldsymbol{e}^{FA}_{\ell}$ are not aligned with the true gradients, $\boldsymbol{e}^{BP}_{\ell}$, computed by backpropagation at this stage of training.

These limitations indicate that the backward flow of information through fixed feedback is insufficient for online training in deeper models. This paper investigates modified plasticity rules to improve the trained model's performance. To that end, a meta-learning framework is adapted to explore a parameterized space of the plasticity rules. 

\subsection{A \reva{meta-learning} approach for discovering interpretable plasticity rules. \label{sec:meta_plasticity}}

Meta-learning is a machine learning paradigm that aims to learn elements of a learning \reva{procedure. 
This} framework consists of a two-level learning scheme: An inner adaptation loop that learns parameters $\boldsymbol{W}$ of a model $f_{\boldsymbol{W}}$ using a parameterized plasticity rule $\mathcal{F}(\boldsymbol{\Theta})$ and an outer meta-optimization loop that modifies the plasticity meta-parameters $\boldsymbol{\Theta}$. The meta-training dataset contains a set of tasks $\{\mathcal{T}_{\revb{\varepsilon}}\}_{\revb{0\leq\varepsilon\leq\mathcal{E}}}$, each consisting of $K$ training data $(\boldsymbol{X}_{train}, \boldsymbol{Y}_{train})$ and $Q$ query data $(\boldsymbol{X}_{query}, \boldsymbol{Y}_{query})$ per class. The former is used to train the model $f_{\boldsymbol{W}}$ while the latter optimizes the meta-parameters $\boldsymbol{\Theta}$. Algorithm~\ref{alg:meta_plast} details the \reva{meta-learning} framework presented in this work.

\begin{algorithm}[H]
	\caption{\reva{meta-learning} algorithm. \label{alg:meta_plast}}
	\begin{algorithmic}
    \STATE Input meta-training set $\{\mathcal{T}_{\revb{\varepsilon}}\}_{\revb{0\leq\varepsilon\leq\mathcal{E}}}=\{(\boldsymbol{X}_{train}^{\revb{\varepsilon}}, \boldsymbol{Y}_{train}^{\revb{\varepsilon}}), (\boldsymbol{X}_{query}^{\revb{\varepsilon}}, \boldsymbol{Y}_{query}^{\revb{\varepsilon}})\}_{\revb{0\leq\varepsilon\leq\mathcal{E}}}$, plasticity rule $\mathcal{F}$, number of episodes $\mathcal{E}$, meta-learning rate $\eta$, and regularization coefficient $\lambda$.
    \STATE Initialize learning parameters $\boldsymbol{\Theta}^{(0)}$.
    \FOR{$\varepsilon=0,\dots,\mathcal{E}$}
        \STATE Initialize network parameters $\boldsymbol{W}^{(0)}$ and $\boldsymbol{B}$.
        \FOR{$(\boldsymbol{x}_{train}^{(i)}, y_{train}^{(i)})\in (\boldsymbol{X}_{train}^{\revb{\varepsilon}}, \boldsymbol{Y}_{train}^{\revb{\varepsilon}})$}
            \STATE Set $\boldsymbol{y}_0=\boldsymbol{x}_{train}^{(i)}$
            \FOR{$\ell=1,\dots,L$}
                \STATE Compute $\boldsymbol{z}_{\ell}$ (Eq.~\ref{eq:z}).
                \STATE Compute $\boldsymbol{y}_{\ell}$ (Eq.~\ref{eq:y}).
            \ENDFOR
            \STATE Compute $\mathcal{L}(\boldsymbol{y}_{L}, y_{train}^{(i)})$.
            \STATE Compute $\boldsymbol{e}_{L}\revb{=\partial \mathcal L/\partial \boldsymbol{z}_L}$.
            \FOR{$\ell=L,\dots,1$}
                \STATE Compute $\boldsymbol{e}_{\ell-1}=\boldsymbol{B}_{\ell,\ell-1}\boldsymbol{e}_{\ell}\odot\sigma'(\boldsymbol{z}_{\ell-1})$ (Eq.~\ref{eq:e_l_backprop}).
                \STATE Update $\boldsymbol{W}_{\ell-1,\ell}^{(i+1)}=\boldsymbol{W}_{\ell-1,\ell}^{(i)} +\mathcal{F} (\boldsymbol{e}_{\ell-1}, \boldsymbol{y}_{\ell-1}, \boldsymbol{e}_{\ell}, \boldsymbol{y}_{\ell}, \boldsymbol{W}_{\ell-1,\ell}^{(i)};\boldsymbol{\Theta}^{(\varepsilon)} )$.
            \ENDFOR
        \ENDFOR
        \STATE Update meta-parameters $\boldsymbol{\Theta}^{(\varepsilon+1)}=\boldsymbol{\Theta}^{(\varepsilon)}-\eta\nabla_{\boldsymbol{\Theta}^{(\varepsilon)}}\left[\mathcal{L}(f(\boldsymbol{X}_{query}^{\revb{\varepsilon}}; \boldsymbol{W}^{(i+1)}), \boldsymbol{Y}_{query}^{\revb{\varepsilon}}) + \lambda\|\boldsymbol{\Theta}^{(\varepsilon)}\|_1\right]$.
    \ENDFOR
	\end{algorithmic}
\end{algorithm}

In each meta-iteration, also known as an episode, a randomly initialized model $f_{\boldsymbol{W}}$ is trained on an online training data sequence. In other words, each adaptation iteration uses a single data point $(\boldsymbol{x}_{train}, y_{train})$ to update $\boldsymbol{W}$. It is worth emphasizing that reinitializing weights $\boldsymbol{W}$ at each episode removes the learning rule's dependence on the weight initialization. The meta-learned plasticity rules are therefore optimized to learn a task starting from a randomly initialized weight matrix. In contrast, meta-optimizing initial weights will adapt meta-parameters $\boldsymbol{\Theta}$ to the later stages of learning, which does not extrapolate to the training lifetime anymore. Moreover, when meta-learning a weight initialization in conjunction with a plasticity rule \reva{({\it e.g.},~\cite{lindsey2020learning})}, it is not clear to what extent improvements in learning can be attributed to the weight initialization versus the meta-learned plasticity rule itself.

\begin{figure}[H]
    \centering
    \includegraphics[width=.95\textwidth]{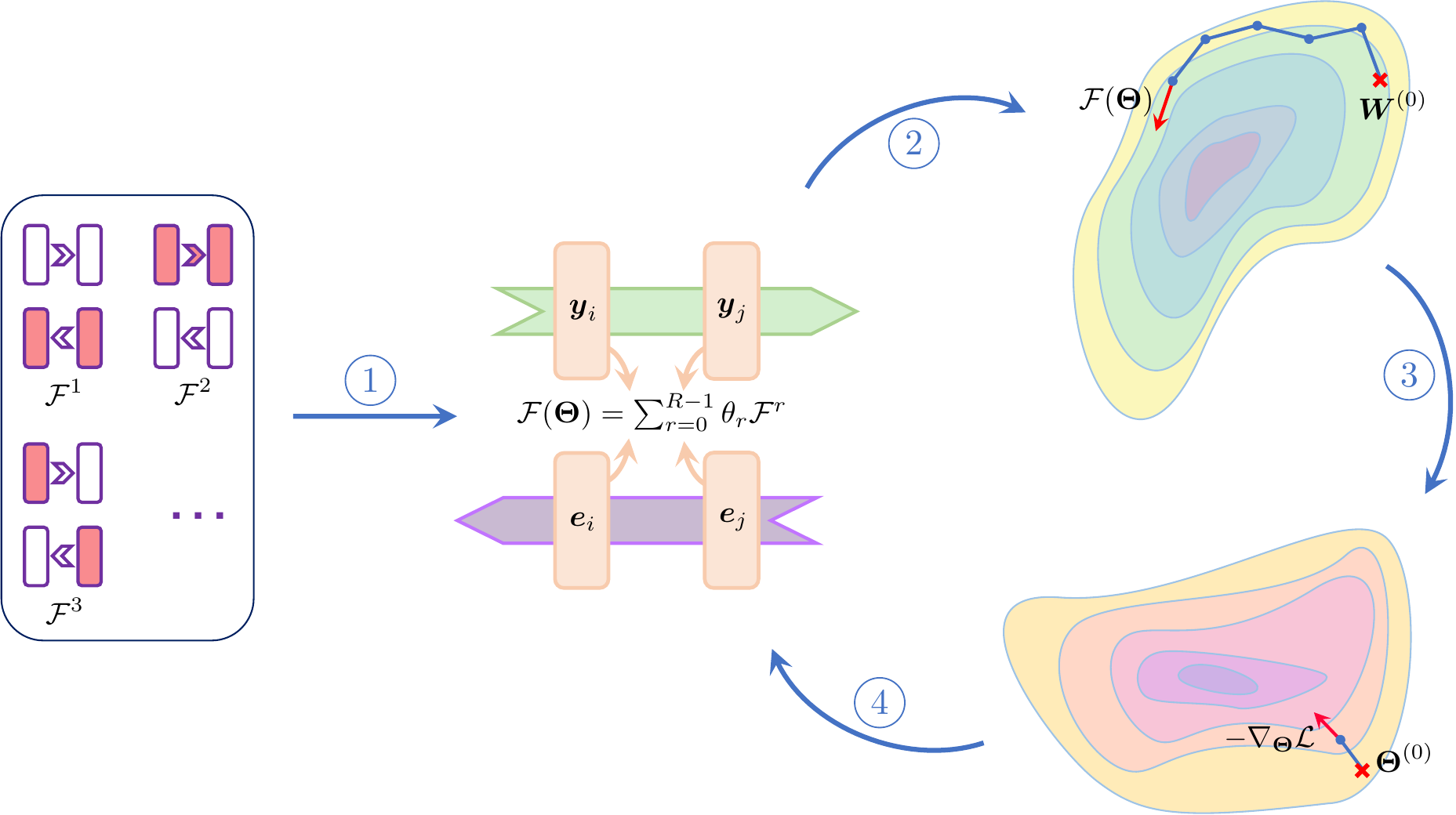}
    \caption{\textbf{Schematic depiction of the \reva{meta-learning} workflow:} \footnotesize \textit{(1) A pool of $R$ biologically \revb{plausible} plasticity terms $\{\mathcal{F}^{r}\}_{0\leq r\leq R-1}$ is exploited to define a plasticity rule $\mathcal{F}(\boldsymbol{\Theta})$ that governs the weight updates of the model $f_{\boldsymbol{W}}$. Each term $\mathcal{F}^{r}$ integrates local elements available to the weight, including pre-synaptic activation $\boldsymbol{y}_{i}$, post-synaptic activation $\boldsymbol{y}_{j}$, pre-synaptic error $\boldsymbol{e}_{i}$, post-synaptic error $\boldsymbol{e}_{j}$, and the current state of the weight $\boldsymbol{W}_{i,j}$. \revb{Such terms are consistent with local plasticity if $\boldsymbol{y}_{i}$ and $\boldsymbol{e}_{i}$ are encoded by the same neuron (see Discussion).}
    The linear combination of these terms defines plasticity rule $\mathcal{F}(\boldsymbol{\Theta})$, where $\boldsymbol{\Theta}=\{\theta_r|0\leq r\leq R-1\}$ is the set of meta-parameters shared across the network. (2) The parameterized local learning rule $\mathcal{F}(\boldsymbol{\Theta})$ is used to navigate the weight parameter space. At each episode $\varepsilon$, $\mathcal{F}(\boldsymbol{\Theta}^{(\varepsilon)})$ iteratively searches for optimized $\boldsymbol{W}$ starting from a random weight $\boldsymbol{W}^{(0)}$. A single data point sampled from $\mathcal{T}_{\revb{\varepsilon}}$'s train set is used at each adaptation step for online training of the model. (3) In the meta-optimization phase, the solution $\boldsymbol{W}$ of the inner loop is used to compute the loss on the query set of task $\mathcal{T}_{\revb{\varepsilon}}$. Then, a gradient-based strategy explores the meta-parameter space to optimize the plasticity meta-parameters $\boldsymbol{\Theta}^{(\varepsilon)}$. (4) The plasticity rule $\mathcal{F}(\boldsymbol{\Theta})$ is reconstructed using the updated meta-parameters $\boldsymbol{\Theta}^{(\varepsilon+1)}$ to guide the weight optimization in the next episode. This procedure is repeated until the meta-parameters converge. In the initial episodes, the unoptimized $\mathcal{F}(\boldsymbol{\Theta})$ is unlikely to direct $\boldsymbol{W}$ to a solution. However, as $\boldsymbol{\Theta}$ converges, $\mathcal{F}(\boldsymbol{\Theta})$ discovers a new direction that may only partially adhere to the direction of the gradient. \label{fig:infographic}}}
\end{figure}

Each episode $\varepsilon$ follows two objectives. The first is to quantify \reva{the} model parameters $\boldsymbol{W}$ using a loss function $\mathcal{L}$, iteratively, on each data point sampled from task $\mathcal{T}_{\revb{\varepsilon}}$'s training set. \reva{Then, given a set of $R$ candidate terms $\{\mathcal{F}^r\}_{0\leq r\leq R-1}$, a parametrized plasticity rule is defined as a linear combination of individual plasticity terms,
\begin{equation}
    \mathcal{F} (\boldsymbol{\Theta}) = \sum_{r=0}^{R-1} \theta_r \mathcal{F}^{r}.
    \label{eq:linear}
\end{equation}
where $\boldsymbol{\Theta}=\{{\theta}_r|0\leq r\leq R-1\}$ is the set of learning parameters shared across layers.
This rule is used to update forward weights, $\boldsymbol{W}$, in the network.} The second objective, dubbed meta-loss, assesses the meta-parameters $\boldsymbol{\Theta}$ by evaluating the loss function $\mathcal{L}$ on the query set of the same task $\mathcal{T}_{\revb{\varepsilon}}$ using the updated model $f_{\boldsymbol{W}}$. \reva{While meta-learning over the pool of plasticity terms $\mathcal{F}(\boldsymbol{\Theta})$ yields an optimized set of meta-parameters, $\boldsymbol{\Theta}$, the resulting plasticity rule consists of too many terms which are difficult to interpret and understand and whose underlying mechanisms may overlap. Therefore, following Occam's razor, we introduce an $L1$ penalty on plasticity coefficients to select for a sparser set of plasticity terms. Mathematically put, the meta-loss is defined as}
\begin{equation}
    \mathcal{L}_{meta}\reva{(\boldsymbol{\Theta})}=\mathcal{L}(f_{\boldsymbol{W}}(\boldsymbol{X}_{query}), \boldsymbol{Y}_{query}) + \lambda \|\boldsymbol{\Theta}\|_1,
    \label{eq:meta_L}
\end{equation}
where $f_{\boldsymbol{W}}$ is the model updated in the adaptation loop and $\lambda$ is a predefined hyperparameter. The regularization term in Eq.~\ref{eq:meta_L} is the $L1$ norm of the meta-parameters, leading the algorithm to favor simplicity in the plasticity model. While weights $\boldsymbol{W}$ are optimized using $\mathcal{F}(\boldsymbol{\Theta})$, meta-parameters $\boldsymbol{\Theta}$ are updated by a gradient-based approach. Figure~\ref{fig:infographic} summarizes the problem's configuration.

\subsubsection{Meta-learning the learning coefficients via backprop and feedback alignment establishes a benchmark \label{sec:F_pseudo}}

Before introducing new plasticity rules, it is necessary to establish the baseline performance for the current learning models for the learning task considered here. To this end, we use the \reva{meta-learning} framework to optimize the learning rate, $\theta$, in Eq.~\ref{eq:F_grad} for backprop and feedback alignment. Since, in these examples, the \reva{meta-learning} model seeks to optimize the meta-parameter rather than selecting one term over the other, the regularization coefficient $\lambda$ in Eq.~\ref{eq:meta_L} is set to zero.

\begin{figure}[H]
    \centering
    \includegraphics[width=.97\textwidth]{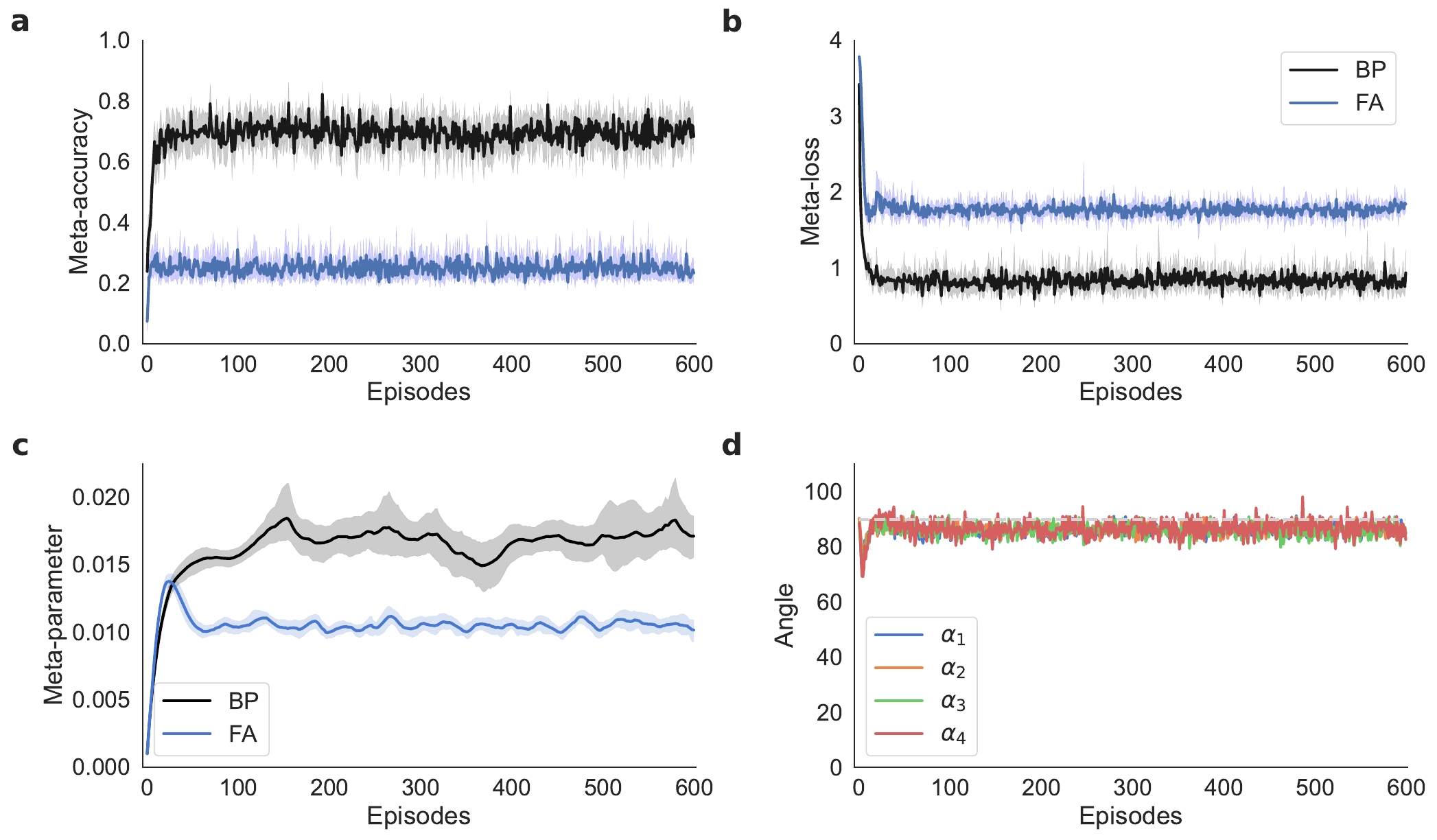}
    \caption{\textbf{Meta-learning coefficients for feedback alignment and backprop.} \footnotesize \textit{(a) Meta-accuracy of feedback alignment (FA) compared to backprop (BP) trained using the \reva{meta-learning} framework (Alg.~\ref{alg:meta_plast}) during \revb{$600$} meta-optimization episodes and (b) the corresponding meta-loss, (c) evolution of the learning rate meta-parameter (initialized to $10^{-3}$) with feedback alignment (FA) compared to backprop (BP) during \revb{$600$} meta-optimization episodes. In this figure, each meta-parameter was optimized separately in a single-parameter meta-optimization problem and is superimposed for comparison. (\nst{d}) Alignment angle $\alpha_{\ell}$ between modulating signals of the feedback alignment $\boldsymbol{e}_{\ell}^{FA}$ and backprop $\boldsymbol{e}_{\ell}^{BP}$ for $l=1$, 2, 3, and 4. For both approaches, $\boldsymbol{e}_5$ is computed using \revb{$\partial \mathcal L/ \partial \boldsymbol{z}_L$} and has the same value, resulting in $\alpha_5=0$\revc{.} }\label{fig:F_pseudo}}
\end{figure}

Figures~\ref{fig:F_pseudo}a~-~\ref{fig:F_pseudo}c compare the performance of the two plasticity rules over \revb{$600$} episodes. First, the reinitialized models $f_{\boldsymbol{W}}$ are trained at each episode using an online stream of $M\times K=250$ data points. Then, the meta-accuracy and meta-loss are evaluated with the query data. Tracing the evolution of the plasticity coefficients in Fig.~\ref{fig:F_pseudo}c shows that the \reva{meta-learning} model converges after $\sim100$ episodes. After convergence, the model trained with feedback alignment is, on average, about $25\%$ accurate in its predictions, whereas the model backpropagated via symmetric feedbacks reaches an approximate accuracy of about $70\%$ \revb{(Fig.~\ref{fig:F_pseudo}a). In addition, the backpropagated model reaches considerably lower loss values as shown in Fig.~\ref{fig:F_pseudo}b.} The comparison shows that the former is not adequately trained with an online data stream in the small data regime. This outcome is further supported by Fig.~\ref{fig:F_pseudo}d, which illustrates the poor alignment of the modulating signals in feedback alignment with the backprop analogs. 

\subsubsection{Biologically \revb{plausible} plasticity rules \label{subsec:F_bio}}

The analysis in section~\ref{sec:F_pseudo} indicated a substantial performance gap between the backprop model and the pseudo-gradient rule with random feedback pathways early in the learning process. However, with the interrupted backward flow as the only distinction between the two rules, the error in the last layer and activations still maintain proper information. Intuitively, introducing new local combinations of these terms to the plasticity rule may restore information flow and improve performance. To that end, we define a set of candidate plasticity terms and use \reva{meta-learning} to uncover combinations that enhance learning. \reva{Meta-learning} helps in two ways: finding the optimized set of meta-parameters for the linear combination of candidate terms and selecting the dominant plasticity terms. While the former avoids cumbersome hand-tuning of the coefficients, the latter provides a tool for systematically studying the space of learning rules.

\begin{figure}[H]
    \centering
    \includegraphics[width=.97\textwidth]{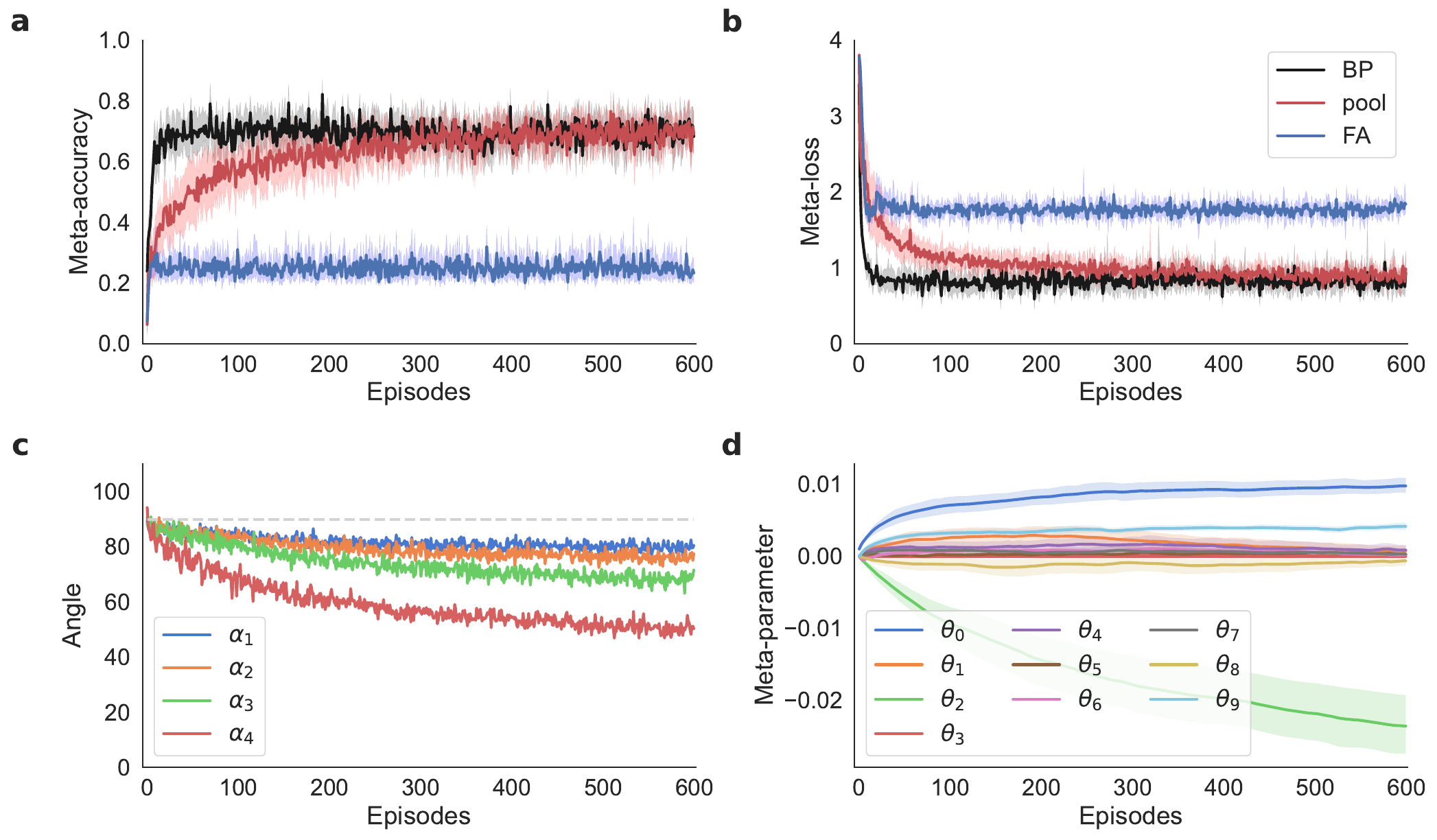}
    \caption{\textbf{Performance of the model trained with the pool of biologically \revb{plausible} plasticity rules $\mathcal{F}^{pool}$:} \footnotesize\textit{(a) Accuracy and (b) loss for $\mathcal{F}^{pool}$ compared to $\mathcal{F}^{0}$ via feedback alignment (FA) and backprop (BP), (c) alignment of the teaching signals of $\mathcal{F}^{pool}$ with the ones for backprop, and (d) convergence of the plasticity coefficients. \label{fig:F_pool}}}
\end{figure}

\reva{

We began by examining a set of $R=10$ plasticity terms and combined them according to Eq.~\ref{eq:linear} to form the learning rule $\mathcal{F}^{pool}$ (see Methods and below for definitions of these rules). Figure}~\ref{fig:F_pool}a--c illustrates the performance of the model. \revb{We set the initial values of the meta-parameters $\{\theta_r\}_{1\leq r<R}$ to 0. As seen in Fig.~\ref{fig:F_pool}a, the model's accuracy initially resembles that of the FA model, but as the meta-optimization continues, the accuracy improves, starting around $10$ episodes. By about 300 meta-iterations, the accuracy approaches that of the BP model. This trend is also echoed in Fig.~\ref{fig:F_pool}b, where the loss initially follows that of the FA learning model but then declines and eventually becomes similar to that of the BP method. In Fig.~\ref{fig:F_pool}c, it is demonstrated that the alignment angles of the teaching signals with their BP counterparts are improved compared to the FA model, seen in Fig.~\ref{fig:F_pseudo}d.} Figure~\ref{fig:F_pool}d shows that the coefficients for all but $3$ terms converge toward zero after about 600 episodes. Those three terms are a pseudo-gradient rule ($\mathcal{F}^{0}$), a Hebbian-like plasticity rule ($\mathcal{F}^{2}$), and Oja's rule ($\mathcal{F}^{9}$). Selecting these three terms and omitting the others gives a simpler plasticity rule of the form
\begin{align}
    \begin{split}
    \mathcal{F}^{bio}(\boldsymbol{\Theta})
     =&-\theta_0 \boldsymbol{e}_{\ell}\boldsymbol{y}_{\ell-1}^T\\
     &-\theta_2 \boldsymbol{e}_{\ell}\boldsymbol{e}_{\ell-1}^T\\
    &\reva{+}\theta_9 (\boldsymbol{y}_{\ell}\boldsymbol{y}_{\ell-1}^T - (\boldsymbol{y}_{\ell}\boldsymbol{y}_{\ell}^T) \boldsymbol{W}_{\ell-1,\ell}),
    \end{split}
    \label{eq:F_bio}
\end{align}
where $\boldsymbol{\Theta}=\{\theta_0, \theta_2, \theta_9\}$ is the set of plasticity meta-parameters. $\mathcal{F}^{bio}$ performs similar to the $\mathcal{F}^{pool}$ (see Supplementary Fig.~\ref{fig:F_bio}) and significantly improves the performance of the feedback alignment method in the low data regime (Fig.~\ref{fig:benchmarks}).

While the \reva{meta-learning} successfully discovers $\mathcal{F}^{bio}$, it is important to interpret the plasticity rule and understand how it leads to improved learning. $\mathcal{F}^{bio}$ consists of three components: a pseudo-gradient term, a Hebbian-style error term, and Oja's rule. In what follows, we study the latter terms separately with the pseudo-gradient term to unveil the underlying reason behind their performance.

\subsubsection*{Hebbian-style error term}

Motivated to understand the Hebbian-style error-based learning term in Eq.~\ref{eq:F_bio}, we rerun the model using a plasticity rule that only includes the modified Hebbian term and the pseudo-gradient term, but omits the third term
\begin{equation}
    \mathcal{F}^{eHebb} ( \boldsymbol{\Theta} ) = -\theta_0\boldsymbol{e}_{\ell}\boldsymbol{y}_{\ell-1}^T-\theta_2\boldsymbol{e}_{\ell}\boldsymbol{e}_{\ell-1}^T.
    \label{eq:F_eHebb}
\end{equation} 

\begin{figure}[H]
    \centering
    \includegraphics[width=.97\textwidth]{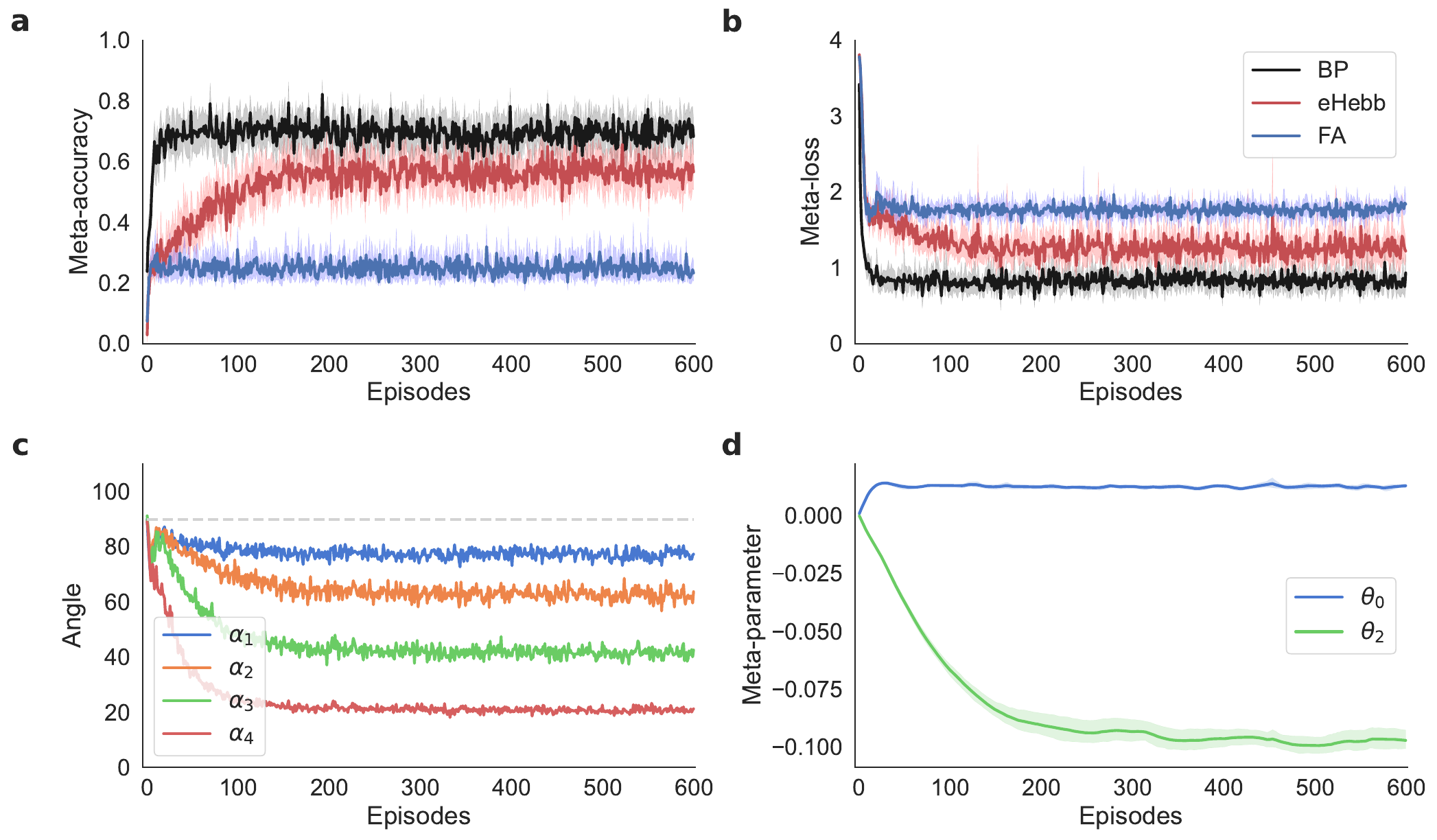}
    \caption{\textbf{Performance of the image classification network trained with $\mathcal{F}^{eHebb}$ (Eq.~\ref{eq:F_eHebb}):} \footnotesize\textit{(a) Meta-accuracy and (b) meta-loss plots for $\mathcal{F}^{eHebb}$ compared to $\mathcal{F}^{0}$ via feedback alignment (FA) and backprop (BP), (c) alignment angles for modulating signals across the network, and (d) convergence of the plasticity coefficients using the \reva{meta-learning} model.} \label{fig:F_eHebb}}
\end{figure}

In Fig.~\ref{fig:F_eHebb}, the \reva{meta-learning} algorithm is used to optimize the coefficients $\theta_0$ and $\theta_{2}$, which are initialized to $10^{-3}$ and zero, respectively. Comparing the accuracy and the loss plot to $\mathcal{F}^{bio}$'s performance (Fig.~\ref{fig:F_bio}) shows that while $\mathcal{F}^{eHebb}$ demonstrates a significant improvement over $\mathcal{F}^{0}$ via feedback alignment, it is yet to reach that of $\mathcal{F}^{bio}$. Despite this, the \reva{teaching signals of $\mathcal{F}^{eHebb}$ are better aligned with the backprop direction than $\mathcal{F}^{bio}$'s} (Fig.~\ref{fig:F_bio}), which indicates that the Hebbian error term is the driving force behind aligning the teaching signals in $\mathcal{F}^{bio}$.

\begin{figure}[H]
    \centering
    \includegraphics[width=.739\textwidth]{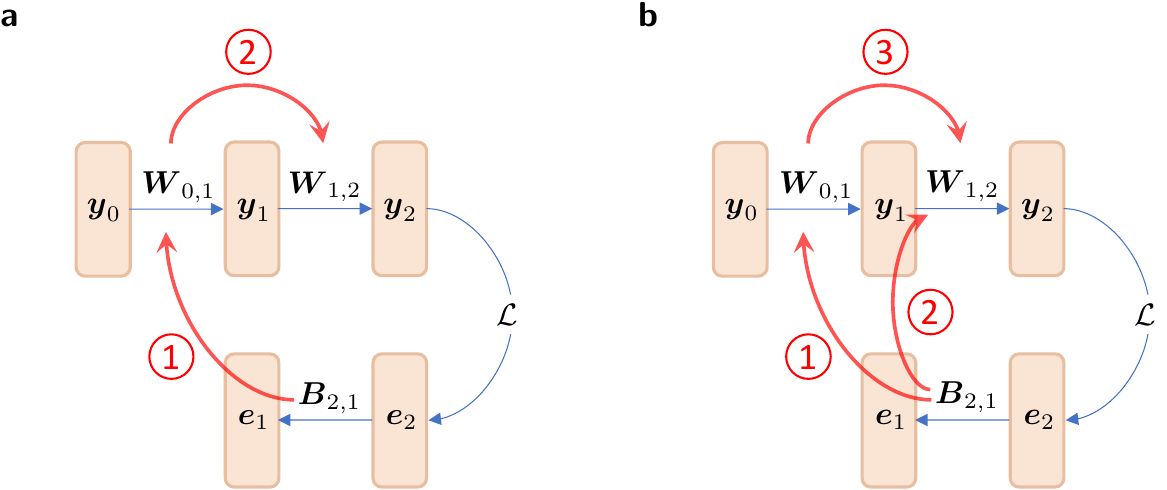}
    \caption{\textbf{Information flow between the forward and backward pathways:} \footnotesize\textit{(a) Both layers are trained with \reva{the rule $\mathcal{F}(\boldsymbol{\Theta})=\theta_0\mathcal{F}^{0}$ via feedback alignment. In this case, information from $\boldsymbol{B}_{2,1}$ is transmitted to $\boldsymbol{W}_{0,1}$ through $\mathcal{F}^{0}$ (\emph{\textcircled{\scriptsize 1}}) and then propagated forward to $\boldsymbol{W}_{1,2}$ (\emph{\textcircled{\scriptsize 2}}).} (b) The first layer is updated with \reva{the rule $\mathcal{F}(\boldsymbol{\Theta})=\theta_0\mathcal{F}^{0}$} via feedback alignment, while the second layer uses \reva{$\mathcal{F}^{eHebb}(\boldsymbol{\Theta})=\theta_0\mathcal{F}^{0}+\theta_2\mathcal{F}^{2}$. Using $\mathcal{F}^{0}$, information from $\boldsymbol{B}_{2,1}$ is communicated to $\boldsymbol{W}_{1,2}$ (\emph{\textcircled{\scriptsize 1}},\emph{\textcircled{\scriptsize 3}}); meanwhile, the presence of $\mathcal{F}^{2}$ sets up a new channel to directly communicate information from $\boldsymbol{B}_{2,1}$ to $\boldsymbol{W}_{1,2}$ (\emph{\textcircled{\scriptsize 2}}).} The blue arrows depict information propagation through the forward and backward paths. The communications between feedback and feedforward pathways are represented with red arrows.} \label{fig:F_eHebb_flow}}
\end{figure}

Figure~\ref{fig:F_eHebb_flow} illustrates how $\mathcal{F}^{eHebb}$ alters the communications between the backward and forward pathways. The \reva{diagram in} Fig.~\ref{fig:F_eHebb_flow}a shows a model solely trained with the $\mathcal{F}^{0}$ via feedback alignment. In this scenario, the information from $\boldsymbol{B}_{2,1}$ flows to $\boldsymbol{W}_{0,1}$ through Eq.~\ref{eq:e_l_backprop}, which is then propagated to $\boldsymbol{W}_{1,2}$ \reva{after} the forward pass. This configuration updates $\boldsymbol{W}_{1,2}$ to align the modulator vector $\boldsymbol{e}_1$ with the backprop counterpart. Nonetheless, this machinery does not sufficiently align the modulating signals when applied to deeper networks with fewer training iterations. In the diagram on the right, the last layer is updated with an additional Hebbian-style plasticity term $\mathcal{F}^{2}$, while the first layer is trained with vanilla $\mathcal{F}^{0}$ rule via feedback alignment. Once again, information from $\boldsymbol{B}_{2,1}$ flows into $\boldsymbol{W}_{0,1}$. However, this time, $\mathcal{F}^{eHebb}$ introduces an auxiliary channel to flow the information from $\boldsymbol{B}_{2,1}$ to $\boldsymbol{W}_{1,2}$. Finally, the forward propagation through the network implicitly transmits the information from $\boldsymbol{B}_{2,1}$ to $\boldsymbol{W}_{1,2}$. The modified rule $\mathcal{F}^{eHebb}$ establishes an explicit supplementary means to communicate between $\boldsymbol{B}_{2,1}$ and $\boldsymbol{W}_{1,2}$, boosts the alignment of $\boldsymbol{e}_1$, and improves the model's performance. \reva{Note that the mechanism in $\mathcal{F}^{0}$ needs two learning iterations to transmit information from $\boldsymbol{B}_{2,1}$ to $\boldsymbol{W}_{1,2}$; information from $\boldsymbol{W}_{0,1}$ propagates to $\boldsymbol{W}_{1,2}$ only after $\boldsymbol{y}_{1}$ is computed with the updated $\boldsymbol{W}_{0,1}$. Meanwhile, $\mathcal{F}^{2}$ does this in the same iteration, carrying out expedited learning.}

To corroborate the argument above, we consider a $3-$layer network trained with $\mathcal{F}^{0}$ rule via feedback alignment and inspect the effect of adding \reva{the error-based Hebbian-style plasticity term $\mathcal{F}^{2}$} on the alignment angles in different layers. To that end, rather than sharing the same learning rule across the network, each layer is updated using one of the $\mathcal{F}^{0}$ rule via feedback alignment or $\mathcal{F}^{eHebb}$ rules. Table~\ref{tab:F_eHebb_alignment} determines that adding the Hebbian error term to the weight update reduces the alignment angle \reva{$\alpha$} between the pre-synaptic error and its backprop analog. A more detailed discussion can be found in Supplementary Notes.

\begin{table}[H]
  \begin{center}
    \begin{tabular}{ccccc}
    \hline
        $\mathcal{F}^{0}$                                                 & $\mathcal{F}^{eHebb}$                                                  & $\alpha_0$ & $\alpha_1$ & $\alpha_2$ \\ \hline
        $\boldsymbol{W}_{0,1}$, $\boldsymbol{W}_{1,2}$, $\boldsymbol{W}_{2,3}$ & -                                                                      & 89.89	& 76.69	& 82.04      \\
        $\boldsymbol{W}_{0,1}$, $\boldsymbol{W}_{2,3}$                         & $\boldsymbol{W}_{1,2}$                                                 & 89.95	& 59.95	& 72.14	     \\
        $\boldsymbol{W}_{0,1}$, $\boldsymbol{W}_{1,2}$                         & $\boldsymbol{W}_{2,3}$                                                 & 90.03	& 75.18	& 29.02	      \\
        $\boldsymbol{W}_{2,3}$                                                 & $\boldsymbol{W}_{0,1}$, $\boldsymbol{W}_{1,2}$                         & 75.29	& 61.23	& 72.56	    \\
        $\boldsymbol{W}_{0,1}$                                                 & $\boldsymbol{W}_{1,2}$, $\boldsymbol{W}_{2,3}$                         & 90.2	& 49.4	& 27.9	      \\
        $\boldsymbol{W}_{1,2}$                                                 & $\boldsymbol{W}_{0,1}$, $\boldsymbol{W}_{2,3}$                         & 84.86	& 74.25	& 30.33	     \\
        -                                                                      & $\boldsymbol{W}_{0,1}$, $\boldsymbol{W}_{1,2}$, $\boldsymbol{W}_{2,3}$ & 77.93	& 49.93	& 28.4	   \\ \hline
    \end{tabular}
  \end{center}
    \caption{\textbf{Effect of the Hebbian-like error learning rule $\mathcal{F}^{eHebb}$ on the alignment of the modulating signals $\alpha_{\ell}$ for different layers:} \footnotesize\textit{The leftmost column includes the parameters updated using $\mathcal{F}^{0}$ with feedback alignment, and the next column indicates layers trained with $\mathcal{F}^{eHebb}$ (Eq.~\ref{eq:F_eHebb}). Angles $\alpha_{\ell}$ \reva{represent the alignment between the modulatory signal $\boldsymbol{e}_{\ell}$ and the backpropagated counterpart $\boldsymbol{e}_{\ell}^{BP}$ at each layer (in degrees)}. Since $\boldsymbol{e}_{0}$ is a synthetic error, the effect of the $\mathcal{F}^{eHebb}$ on $\boldsymbol{W}_{0,1}$ alone has been excluded. The model is trained for $500$ episodes, and the computed angles are averaged after a burn-in period of $100$ episodes. \label{tab:F_eHebb_alignment}}}  
\end{table}
For a more precise, mathematical intuition of the effects that $\mathcal{F}^{eHebb}$ has on weights, we show in Supplementary Notes that, in a linear network model under reasonable approximating assumptions,  
\[
\mathbb E\left[\left.\boldsymbol{e}_\ell \boldsymbol{e}_{\ell-1}^T\, \right|\,\boldsymbol{B}_{\ell,\ell-1}\right]\propto \boldsymbol{B}_{\ell,\ell-1}^T
\]
for layers, $\ell=1,2,\ldots,L-1$.  
Thus, the term $\boldsymbol{e}^T_{\ell}\boldsymbol{e}_{\ell-1}$ in $\mathcal{F}^{eHebb}$ pushes $\boldsymbol{W}_{\ell-1, \ell}$ toward the transpose of $\boldsymbol{B}_{\ell, \ell-1}$, resulting in faster alignment of the modulatory signals with the backprop algorithm's error vectors and more efficient learning.

\subsubsection*{Oja's rule}

Eq.~\ref{eq:F_bio} proposes a plasticity rule to train deep networks using fixed feedback matrices. Above, we demonstrated that the Hebbian-style learning term improves the trained model's performance by improving the modulatory signals' alignments with the backpropagated analogs. Here, we look at the remaining plasticity term in Eq.~\ref{eq:F_bio}: Oja's rule, a purely local learning rule that updates the weights based on its current state and the local activations in the forward path. To this end, we redefine the plasticity rule as a linear combination of the pseudo-gradient term and Oja's rule
\begin{equation}
    \mathcal{F}^{Oja}(\boldsymbol{\Theta}) = -\theta_0\boldsymbol{e}_{\ell}\boldsymbol{y}_{\ell-1}^T \reva{+} \theta_9 (\boldsymbol{y}_{\ell}\boldsymbol{y}_{\ell-1}^T - (\boldsymbol{y}_{\ell}\boldsymbol{y}_{\ell}^T) \boldsymbol{W}_{\ell-1,\ell}).
    \label{eq:F_Oja}
\end{equation}

\begin{figure}[H]
    \centering
    \includegraphics[width=.97\textwidth]{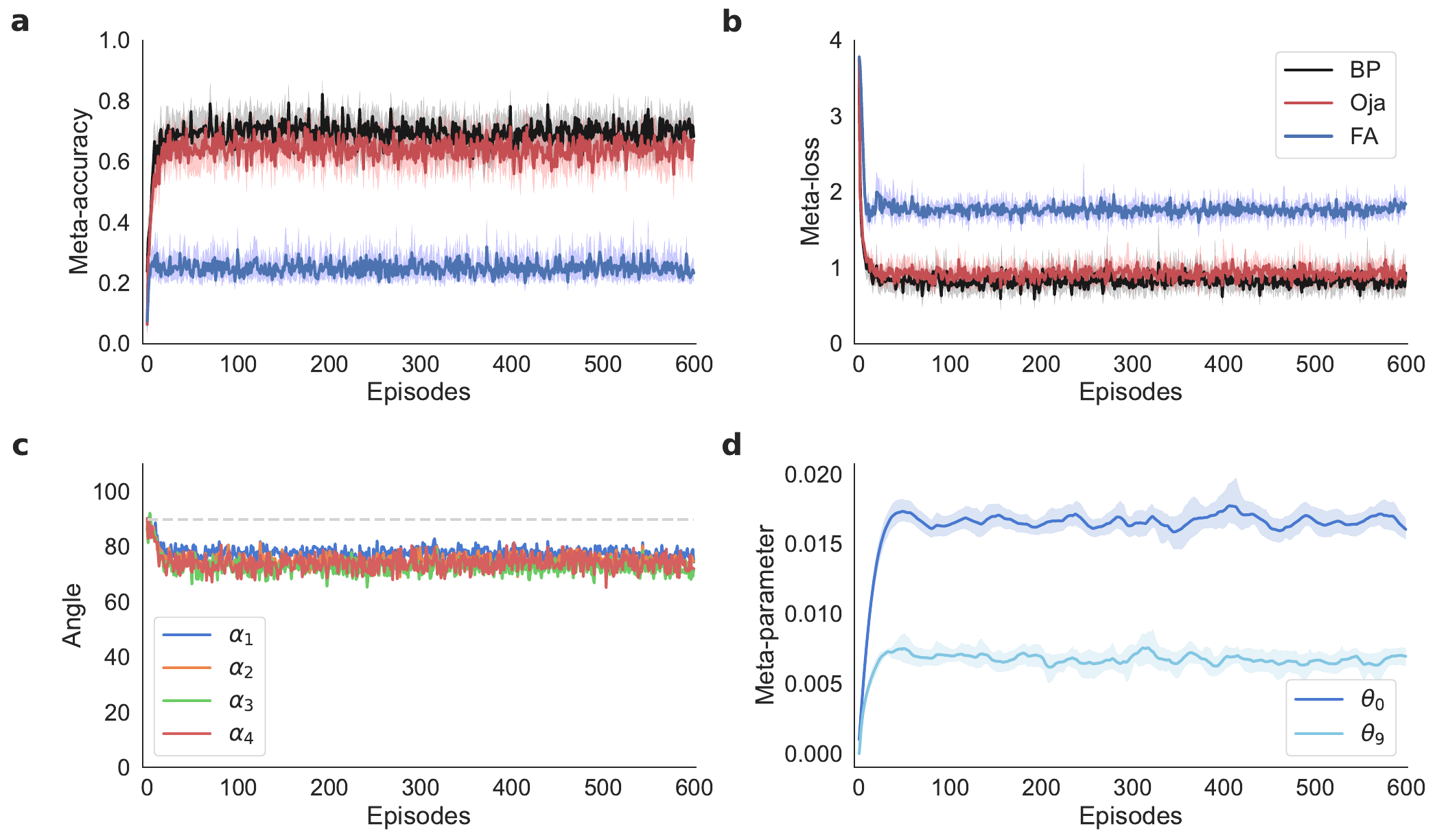}
    \caption{\textbf{Performance of the model trained using $\mathcal{F}^{Oja}$ (Eq.~\ref{eq:F_Oja}) through fixed backward connections:} \footnotesize\textit{(a) Meta-accuracy and (b) meta-loss of $\mathcal{F}^{Oja}$ compared to $\mathcal{F}^{0}$ learning rule via feedback alignment (FA) and backprop (BP), (c) alignment of modulating signals of $\mathcal{F}^{Oja}$ with backprop's teaching signals, and (d) evolution of the plasticity meta-parameters.} \label{fig:F_Oja}}
\end{figure}

We initialize $\theta_0$ to $10^{-3}$ and $\theta_9$ to zero and employ Alg.~\ref{alg:meta_plast} to optimize the set of meta-parameters $\boldsymbol{\Theta}$. Figures~\ref{fig:F_Oja}a and~\ref{fig:F_Oja}b illustrate that adding Oja's rule to the pseudo-gradient term enhances the model's accuracy when backward connections are fixed. Figure~\ref{fig:F_Oja}c presents the angles between the teaching signals ensued by Eq.~\ref{eq:F_Oja} and the corresponding backpropagated ones. While the accuracy and loss are significantly improved, contrary to expectations, Oja's rule does not substantially reduce the alignment angles \reva{(Fig.~\ref{fig:F_Oja}c). In fact, alignment angles are only slightly smaller when using Oja's rule compared to using pure FA, as seen by comparing Fig.~\ref{fig:F_Oja}c to Fig.~\ref{fig:F_pseudo}d. This contrasts with alignment angles for $\mathcal{F}^{eHebb}$ and $\mathcal{F}^{bio}$, which are greatly reduced in deeper layers compared to $\mathcal{F}^{Oja}$ (compare Fig.~\ref{fig:F_Oja}c to Figs.~\ref{fig:F_eHebb}c and \ref{fig:F_bio}c).}

\begin{figure}[H]
    \centering
    \includegraphics[width=.97\textwidth]{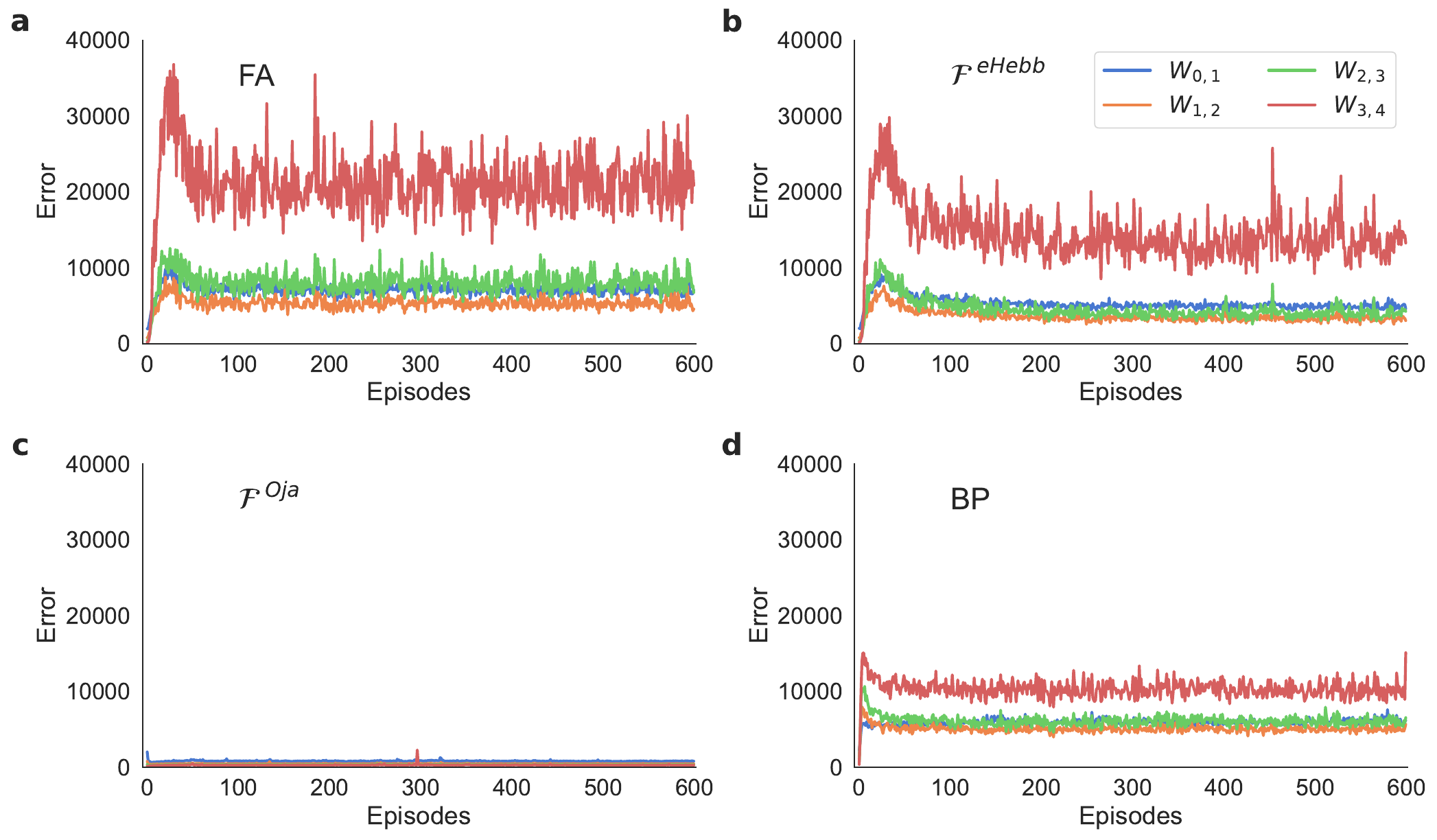}
    \caption{\textbf{Orthonormality error throught a deep network for different plasticity rules:} \footnotesize\textit{Orthonormality errors are measured by Eq.~\ref{eq:orth_err} for different layers of a 5-layer deep network. The model is trained using (a) $\mathcal{F}^{0}$ via feedback alignment (FA), (b) $\mathcal{F}^{eHebb}$, (c) $\mathcal{F}^{Oja}$, and (d) backprop (BP). In this comparison, the last layer has been excluded.} \label{fig:orth}}
\end{figure}

\revc{Inspecting Fig.~\ref{fig:F_Oja} suggests that} rather than helping to align the modulating signals, Oja's rule helps by entirely circumventing the backward path. \revc{Oja's rule implements a Hebbian learning rule subjected to an orthonormality constraint on the weights~\cite{oja1982simplified}. In Eq.~\ref{eq:F_Oja}, $\boldsymbol{y}_{\ell-1}$ and $\boldsymbol{y}_{\ell}$ denote post-nonlinearity activations (as stated in Eq.~\ref{eq:y}), resulting in the $\mathcal{F}^9$ plasticity rule to implement a non-linear version of Oja's rule. When trained iteratively, this non-linear variation implements a recursive non-linear algorithm for Principal Component Analysis~\cite{oja1991data, oja1992principal}. Previous studies on the convergence of Oja's rule have shown that for a compression layer, where $dim(\boldsymbol{y}_{\ell-1})>dim(\boldsymbol{y}_{\ell})$, rows of the weight matrix $\left(\boldsymbol{W}_{\ell-1,\ell}\right)_1, \dots, \left(\boldsymbol{W}_{\ell-1,\ell}\right)_{dim(\boldsymbol{y}_{\ell})}$ will tend to a rotated basis in the $dim(\boldsymbol{y}_{\ell})-$dimensional subspace spanned by the principal directions of the input $\boldsymbol{y}_{\ell-1}$~\cite{williams1985feature}. 

We demonstrate that incorporating Oja's rule into Feedback Alignment improves feature map extraction in the forward path through unsupervised learning, despite $\mathcal{F}^{Oja}$ not recursively applying pure Oja's rule.} By analyzing the continuous-time differential equation corresponding to the Oja's learning rule, Williams~\cite{williams1985feature} and Oja~\cite{oja1992principal} establish the stability limits for this rule. In a compression layer, the fixed point of Oja's rule is a stable solution if $\boldsymbol{W}_{\ell-1,\ell}\boldsymbol{W}^T_{\ell-1,\ell}=\boldsymbol{I}$. This conclusion can be used to derive a proximity measure~\cite{karhunen1994representation, karhunen1995generalizations, karayiannis1996accelerating} of the estimated $\boldsymbol{W}_{\ell-1,\ell}$ to a stable solution of Oja's rule in the presence of non-linear activations. The error
\begin{equation}
    E_{W}=\left\|\boldsymbol{z}_{\ell}-\boldsymbol{W}_{\ell-1, \ell} \bar{\boldsymbol{y}}_{\ell-1}\right\|^{2}_2,
    \label{eq:orth_err}
\end{equation}
where
\begin{equation*}
    \bar{\boldsymbol{y}}_{\ell-1}= \boldsymbol{W}_{\ell-1, \ell}^T\sigma\left(\boldsymbol{z}_{\ell}\right),
\end{equation*}
can define this measure. Figure~\ref{fig:orth} studies this orthonormality measure in models trained with different plasticity rules. Results show that using Oja's rule will render the weight matrices increasingly orthonormal, reducing the correlation in weight rows and improving the feature extraction in these layers. \reva{These findings indicate that introducing Oja's rule alone can help with the problem of slow learning caused by random feedback connections.}

\revc{The architecture of a classifier network includes initial layers that act as feature extractors, creating hidden representations for the final layer. This last layer, dubbed predictor, maps the hidden feature representations to the target class for the given input image. To improve the classifier's performance, a plasticity rule that enhances feature extraction in the earlier layers is beneficial. However, this rule has no grounds to positively impact the predictor layer's performance. Despite this, for comprehensiveness, we also applied the plasticity rule $\mathcal{F}^{Oja}$ to the final layer and found no detrimental effect on the model's performance.}
Ultimately, rather than improving the alignments, \revc{$\mathcal{F}^{Oja}$} provides embeddings that facilitate more effective learning in the last layer.

\section{Discussion}

Despite the dominance of the backpropagation algorithm as the primary technique to train deep neural networks, its biological plausibility remains a significant ground for contest~\cite{whittington2019theories,lillicrap2020backpropagation}. In particular, the presence of feedback synaptic projections that are precisely symmetric to the forward projections is not biologically realistic. Previous work~\cite{lillicrap2016random} showed that learning can be achieved without this symmetry using feedback connections that are randomly sampled, not tied to the forward path, and fixed throughout the training process. While a breakthrough, this method is susceptible to diminished performance when training deeper networks or using smaller batch sizes~\cite{amit2019deep, bartunov2018assessing}. The latter is a challenge for online learning.

A recent body of work attempts to improve learning through asymmetric feedback connections. They either rewire fixed feedback connections, use plastic feedback connections that are updated through an auxiliary plasticity rule, or impose partial symmetry in the backward network~\cite{nokland2016direct, lindsey2020learning, miconi2018backpropamine, miconi2018differentiable, liao2016important}. 
Our work accelerates \revb{the learning process by enhancing the rules that govern neural plasticity while transmitting teaching signals through fixed connections. Our proposed rules for plasticity are based on biologically motivated learning principles, like Oja's rule, or have been inspired by them, such as the error-based Hebbian rule.} A linear combination of these terms yields a parameterized learning rule. To overcome the arduous hand-tuning of these hyper-parameters, we use a \reva{meta-learning} approach that systematically explores the pool of candidate plasticity rules. This approach consists of an inner loop that learns a task and an outer loop that updates the plasticity coefficients. The inner loop always starts from randomly initialized weights, so the model must learn to learn from scratch. Moreover, the inner loop learns from an online stream of training data, simulating real-time learning in the brain. 

\reva{To assure interpretability of our meta-learned learning rule, we expressed the rule as a linear combination of individual plasticity terms, imposed an L1 penalty on the coefficients, and used meta-parameter sharing between all update rules. Many terms in the pool of plasticity rules can be redundant and employ identical or overlapping mechanisms but only differ in their efficiency, \textit{i.e.}, computational cost or the number of required learning iterations to operate. Employing an L1-penalized meta-loss decreases the count of plasticity terms that work in parallel. Additionally, while sharing the same meta-parameters across layers may limit the model's freedom in learning, it is a vital component for discovering a global learning rule, leaving the door open to investigate the revealed terms.}

Using this \reva{meta-learning} approach, we discover two plasticity rules that accelerate learning through fixed feedback connections. 
\revb{The first, an error-based Hebbian rule, combines the errors of pre- and post-synaptic layers to update forward projecting weights. The second rule, known as Oja's rule, combines pre- and post-synaptic activations with the connection's current state to update weights.} We investigated each plasticity rule, its underlying mechanism, and how it contributes to learning, revealing two distinct mechanisms behind them. First, the Hebbian-like error term improves performance by modifying the flow of information through the backward path. It introduces an auxiliary channel to communicate information about the backward connections to the forward weights. As a result, it accelerates learning by better aligning modulating signals with the ones transmitted through a symmetric feedback connection. Ultimately, the modified plasticity alters the training to resemble backpropagation.  Unlike the Hebbian-like rule, Oja's rule does not directly affect the flow of the feedback signals. Instead, it acts only on the forward path, implementing an unsupervised learning scheme that extracts feature maps independently of the labels and loss. The updated weight rows approximate an orthonormal basis in the subspace spanned by PCA eigenvectors of the pre-synaptic activations~\cite{oja1992principal}. The  strengthened signal separation capabilities in the earlier layers improve predictions made by the output layer.

\revb{While synaptic plasticity in the brain is mediated by a vast array of biophysical processes, the changes to a single synaptic weight largely depend on the activity of its pre-synaptic and post-synaptic neurons and the current weight, a property known as ``local'' plasticity. For the plasticity rules used in our study (with the exception of Oja's rule), weight updates depend on activations from a forward pass {\it and} error signals from a backward pass. Since these quantities were used to update the {\it forward} projecting weights, this raises the question of whether the plasticity rules are truly local. The answer to this question depends on the biological interpretation of the forward and backward passes. 

Under one interpretation, separate populations of neurons encode the forward and backward passes, {\it i.e.}, the neurons encoding $\boldsymbol{e}_{\ell}$ are distinct from those encoding $\boldsymbol{y}_\ell$. Under this interpretation, the plasticity rules used in this study are not strictly local. 

Under another interpretation, forward activations and backward errors are represented by the same neural populations, {\it i.e.}, the same neurons encode $\boldsymbol{e}_{\ell}$ and $\boldsymbol{y}_\ell$. Under this interpretation, all of the plasticity rules used in this study are local. 
There are several models for how this multiplexing of forward and backward signals could be achieved (see~\cite{whittington2019theories} for a review). For example, activations and errors could be represented at separate points in time by the same neurons. 

Alternatively, recent work hypothesizes that activations and errors are encoded separately in the basal and apical dendrites of the same cortical pyramidal neurons~\cite{sacramento2018dendritic}.  
Along similar lines, a growing body of work posits that activations and errors are multiplexed by the distinction between bursts and single action potentials, which are communicated separately by synaptic projections onto the soma versus apical dendrites of pyramidal neurons~\cite{kording2001supervised,naud2018sparse,payeur2021burst}.  
The dependence of synaptic plasticity on the morphological site of the synaptic contact and on the type of spiking (bursts versus individual spikes) is well established in experiments~\cite{paulsen2000natural,letzkus2006learning,kampa2006requirement,nevian2006spine,froemke2006contribution}. 
Under these models, established biophysical properties of cortical synapses can produce plasticity rules like ours that multiplex forward and backward propagating information to update weights. 
Networks in~\cite{payeur2021burst} rely on weight decay to approximately align forward and backward weights~\cite{akrout2019deep}, while some networks in~\cite{sacramento2018dendritic} rely on random feedback alignment. Hence, our meta-learned plasticity rules could improve learning in those models.

Our meta-learning approach isolated three plasticity terms: a backprop-like rule ($\mathcal F^0$), Oja's rule~\cite{oja1982simplified} ($\mathcal {F}^9$), and a rule we refer to as eHebb ($\mathcal F^2$). Possible biological implementations of Oja's rule and the backprop-like rule have been studied in great depth in previous work~\cite{sacramento2018dendritic,whittington2019theories,lillicrap2020backpropagation,payeur2021burst}. The eHebb rule could be implemented in a similar way to the backprop-like rule. For example, under the model in~\cite{payeur2021burst}, eHebb would change synaptic weights in response to the co-occurrence of pre- and post-synaptic bursts. Plasticity is strongly mediated by firing rates and intracellular calcium~\cite{graupner2012calcium,graupner2016natural}, both of which are elevated during bursts.}

As the \revb{eHebb's} mechanism tends to align modulating signals with the symmetric counterparts, its performance may at best match that of backprop. However, as  Oja's rule does not aim to imitate backprop, its performance is not bounded by that of backprop, and hence it can also be used to enhance learning in symmetric feedback models. For instance, we realized that adding Oja's plasticity rule to the gradient-based learning term accelerates learning for poorly initialized networks. This observation explains why the improved performance in the fixed feedback model may outperform learning in the symmetric case. A similar concept was used in the earlier works to initialize internal representations of the neural networks~\cite{karayiannis1996accelerating}. However, that work used weights preprocessed by Oja's rule to start gradient-based learning rather than using both terms simultaneously as the plasticity rule. Hence, our results demonstrate the utility of the proposed \reva{meta-learning} approach as a tool for combining different learning terms as a single parameterized learning rule. 

\revb{We used meta-learning to find plasticity rules that can learn effectively under the biologically relevant setting where forward and backward weights are not explicitly aligned. But our meta-learning technique can be applied more broadly to identify plasticity rules that overcome other biological constraints in various contexts and models. For instance, our study only focused on plasticity in forward connections; however, backward projections in the brain can also exhibit plasticity. Our meta-learning approach can be extended to discover plasticity rules for backward connections in such settings.} 
Another interesting future direction is to meta-learn the {\it architecture} of the feedback pathways instead of (or in addition to) the plasticity parameters. That is, to simultaneously provide both direct~\cite{nokland2016direct} and regular~\cite{lillicrap2016random} feedback pathways and allow the meta-learning algorithm to pick the most efficient path to carry the teaching signals to each layer. 

In another direction, our meta-parameter sharing approach could be partially relaxed without learning a new plasticity rule for each connection. For example, one could consider a network with several neural populations and a shared plasticity rule for each pair of populations. This approach could help understand the role of distinct neuron types and populations in biological circuits. 

We focused on meta-learning biologically \revb{plausible} plasticity rules, but our approach can also be applied to discover learning rules that satisfy other constraints or optimize other meta-loss functions. For example, the approach can be used to find learning rules that can be implemented in non-standard hardware like neuromorphic chips or optical networks, or to discover learning rules that minimize energy consumption or other factors. 

In summary, we developed and tested a \reva{meta-learning} approach designed to produce simple, interpretable plasticity rules that can effectively learn on new data. First, using randomly initialized weights on each iteration of the outer loop (instead of meta-learning the initialization) and using online learning in our inner loop encouraged plasticity rules that can perform online learning from scratch. Secondly, meta-parameter sharing yielded a vastly smaller set of learned plasticity rules compared to learning a plasticity rule for each synapse. Finally, an L1 penalty on plasticity coefficients promoted sparsity within the learning rule, ultimately yielding a small set of plasticity terms that are more readily interpreted. Our results demonstrate the utility of this approach for discovering and interpreting plasticity rules. Taken together, our work opens new avenues to the application of \reva{meta-learning} for discovering interpretable plasticity rules that satisfy biological or other constraints.

\clearpage

\section{Methods\label{sec:methods}}

\subsection{Models\label{app:models}}

Figure~\ref{fig:benchmarks} performs a 10-way classification on the MNIST dataset, with images resized to $28\times28$ dimensions. The model is a 5-layer fully connected neural network with dimensions 784-170-130-100-70-47. Hidden layers use the softplus activation function 
\begin{equation}
    \sigma (\boldsymbol{z}_{\ell})=\frac{1}{\beta}\log(1+\exp(\beta\boldsymbol{z}_{\ell})),
    \label{eq:sp}
\end{equation}
with $\beta=10$. The output layer uses the softmax activation function. Figures~\ref{fig:F_pseudo}~-~\ref{fig:F_eHebb} and~\ref{fig:F_Oja}~-~\ref{fig:orth} perform 5-way classification on the EMNIST dataset. These figures use the same architecture as Fig.~\ref{fig:benchmarks}. For Tab.~\ref{tab:F_eHebb_alignment}, the model conducts a 5-way classification on the EMNIST dataset with an image size of $28\times28$. The model is a 3-layer fully connected neural network with dimensions 784-130-70-47. Like the rest of the paper, hidden layers use softplus non-linearity with $\beta=10$, while the output layer uses softmax.

\revb{In the fixed feedback pathway problem, the weights and feedback connections are initially set to random values that differ from each other. Both symmetric and fixed feedback models utilize the Xavier method~\cite{glorot2010understanding} to re-initialize forward and backward connections at the start of each meta-learning episode.}

In Figs.~\ref{fig:F_pool},~\ref{fig:F_eHebb},~\ref{fig:F_Oja}, and~\ref{fig:orth}, and Tab.~\ref{tab:F_eHebb_alignment}, we set the initial value for the learning rate $\theta_0$ of the term $\mathcal{F}^0$ to $10^{-3}$ and set all other hyper-parameters to zero.

\revc{All plots depict the mean outcome over $20$ trials, each with different initial weights and feedback matrices. The shaded region in the loss, accuracy, and meta-parameters plots illustrates the $98\%$ confidence interval, determined through bootstrapping with $500$ samples.}

\subsection{Candidate learning terms \label{app:F_bio}}

Section~\ref{subsec:F_bio} presented a plasticity rule that improves the model's performance in the presence of fixed random feedback connections (Eq.~\ref{eq:F_bio}). We employed the \reva{meta-learning} framework described in section~\ref{sec:meta_plasticity} to explore a set of local learning rules to discover such a plasticity term. This set of terms is defined as
\begin{align*}
    \mathcal{F}^{0} &= \reva{-} \boldsymbol{e}_{\ell} \boldsymbol{y}_{\ell-1}^T, \\
    \mathcal{F}^{1} &= \reva{-} \boldsymbol{y}_{\ell} \boldsymbol{e}_{\ell-1}^T, \\
    \mathcal{F}^{2} &= \reva{-} \boldsymbol{e}_{\ell} \boldsymbol{e}_{\ell-1}^T, \\
    \mathcal{F}^{3} &= \reva{-} \boldsymbol{W}_{\ell-1, \ell}, \\
    \mathcal{F}^{4} &= \reva{-} \boldsymbol{1}_{\ell} \boldsymbol{e}_{\ell-1}^T, \\
    \mathcal{F}^{5} &= \reva{-} \boldsymbol{e}_{\ell}\boldsymbol{1}_{\ell}^T\boldsymbol{y}_{\ell}\boldsymbol{y}_{\ell-1}^T, \\
    \mathcal{F}^{6} &= \reva{-} \boldsymbol{y}_{\ell}\boldsymbol{y}_{\ell}^T\boldsymbol{W}_{\ell-1, \ell}\boldsymbol{e}_{\ell-1}\boldsymbol{e}_{\ell-1}^T, \\
    \mathcal{F}^{7} &= \reva{-} \boldsymbol{e}_{\ell}\boldsymbol{y}_{\ell}^T\boldsymbol{W}_{\ell-1, \ell}\boldsymbol{e}_{\ell-1}\boldsymbol{y}_{\ell-1}^T, \\
    \mathcal{F}^{8} &= \reva{-} \boldsymbol{y}_{\ell}\boldsymbol{y}_{\ell-1}^T\boldsymbol{W}_{\ell-1, \ell}^T\boldsymbol{e}_{\ell}\boldsymbol{e}_{\ell-1}^T, \\
    \mathcal{F}^{9} &= \reva{(}\boldsymbol{y}_{\ell} \boldsymbol{y}_{\ell-1}^T - (\boldsymbol{y}_{\ell} \boldsymbol{y}_{\ell}^T) \boldsymbol{W}_{\ell-1, \ell}).
\end{align*}
\reva{The rules above are local in the sense that the updates to the $j,k$th entry of $\boldsymbol{W}_{\ell-1,\ell}$ depend only on the $k$th entry of $\boldsymbol{e}_{\ell-1}$ and $\boldsymbol{y}_{\ell-1}$, the $j$th entry of $\boldsymbol{y}_{\ell}$ and $\boldsymbol{e}_\ell$, and the $j,k$th entry of $\boldsymbol{W}_{\ell-1,\ell}$. This notion of locality assumes that errors and activations are encoded in the same neurons (see Discussion). 
Even under this constraint of locality, there is an unlimited number of possible plasticity rules to choose from. To form the list above, we first considered all quadratic combinations of activations and errors except we omitted pure Hebbian plasticity ($\boldsymbol{y}_{\ell} \boldsymbol{y}_{\ell-1}^T$) because we found that it leads to unstable network dynamics (a blowup of activations). Instead, we replaced it with Oja's rule $\mathcal F^9$, which adds a stabilizing term onto pure Hebbian plasticity. Additional terms were added to test the viability of higher order plasticity terms.}

Computing the learning terms $\mathcal{F}^1$, $\mathcal{F}^2$, $\mathcal{F}^4$, $\mathcal{F}^6$, $\mathcal{F}^7$, and $\mathcal{F}^8$ requires a pre-synaptic error term. In order to update the weights in the first layer $\boldsymbol{W}_{0,1}$, where there is no pre-synaptic error, we define a synthetic error $\boldsymbol{e}_{0}$ using Eq.~\ref{eq:e_l_backprop} and the activation function in Eq.~\ref{eq:sp}, such that
\begin{equation*}
    \boldsymbol{e}_{0}:=\boldsymbol{B}_{1,0}\boldsymbol{e}_{1} \odot(1-\exp(-\beta\boldsymbol{y}_0)).
\end{equation*}

\subsection{Meta-Training \label{sec:train}}

Section~\ref{sec:meta_plasticity} presented a \reva{meta-learning} framework for swiftly exploring a pool of plasticity terms and uncovering combinations that exceed the performance of the existing plasticity rule. We demonstrate this by training a classifier network, which performs a $5-$way classification on $28\times28$ images. The cross-entropy function evaluates the loss in the adaptation loop, whereas the meta-loss is determined by Eq.~\ref{eq:meta_L}. While, in principle, any optimization algorithm, such as evolutionary methods, can be used to optimize $\boldsymbol{\Theta}$, the algorithm presented in Alg.~\ref{alg:meta_plast} uses ADAM~\cite{kingma2014adam}, a gradient-based optimization technique, with a meta-learning rate of $10^{-3}$. 

In the meta-optimization phase, this gradient-based optimizer differentiates through the unrolled computational graph of the adaptation phase. Thus, the non-linear layers are double differentiated, once to compute $\boldsymbol{e}_L$ and a second time by the meta-optimizer. This arrangement will only allow a two-times differentiable non-linear layer, which prohibits using the Rectified Linear Unit, ReLU, as the activation function $\sigma$. Instead, we use the softplus function (Eq.~\ref{eq:sp}), a continuous, twice-differentiable approximation of the ReLU function. In Eq.~\ref{eq:sp}, parameter $\beta$ controls the smoothness of the function. \reva{Furthermore, the L1 norm used in the meta-loss (Eq.~\ref{eq:meta_L}), defined by the absolute value function, is not continuously differentiable at every point. However, it is commonly used in deep learning in conjunction with stochastic gradient descent (SGD)~\cite{goodfellow2016deep}. In PyTorch and other deep learning frameworks, the derivative of the absolute value function is typically defined as zero at zero.}

In the present examples, each task contains $M=5$ labels. Consequently, assembling a diverse set of $5-$way classification tasks requires a database with a large number of classes. Thus, databases such as MNIST~\cite{lecun1998gradient}, which only has ten classes, are unsuitable for proper meta-training. On the other hand, in each episode, the classifier $f_{\boldsymbol{W}}$ is reinitialized with random weights $\boldsymbol{W}$. Therefore, the task should contain enough data points per class to train $f_{\boldsymbol{W}}$ adequately. Hence, databases such as Omniglot~\cite{lake2015human} with only 20 data points per character designed for few-shot learning ({\it e.g.}, with meta-optimized $\boldsymbol{W}$) are impractical in the present framework. In the current work, meta-training tasks are made from the EMNIST database~\cite{cohen2017emnist}. This database contains 47 classes, making it a good candidate for the \reva{meta-learning} framework. Each task contains $K=50$ training and $Q=10$ query data points per class. 

Notably, the use of $K=50$ training data per class with $M=5$ classes in each episode means that the metalearned plasticity rule needs to train a randomly initialized network with only $250$ training data points. Hence, our models are in a low data regime without the benefit of pre-trained weights that are often used for few-shot learning. 

\nst{
\section{Data Availability\label{sec:data}}

In this study, the EMNIST database~\cite{cohen2017emnist} was used for meta-learning experiments. This database is publicly accessible at https://doi.org/10.1109/IJCNN.2017.7966217. Additional benchmarking was done using the MNIST dataset~\cite{lecun1998gradient} and the FashionMNIST dataset~\cite{xiao2017fashion}. These datasets can be found at http://yann.lecun.com/exdb/mnist and https://github.com/zalandoresearch/fashion-mnist, respectively. Source data are provided with this paper.

\section{Code Availability\label{sec:code}}


The PyTorch-based implementation and script files used to generate the results in this paper will be publicly accessible at https://github.com/NeuralDynamicsAndComputing/ upon publication.
}

\newpage
\bibliographystyle{unsrtFirstInit}
\bibliography{references}

\clearpage
\beginsupplement
\renewcommand{\theequation}{S.\arabic{equation}}
\setcounter{equation}{0}

\section*{Supplementary Material}

\subsection*{Performance of DFA}

\begin{figure}[H]
    \centering
    \includegraphics[width=.475\textwidth]{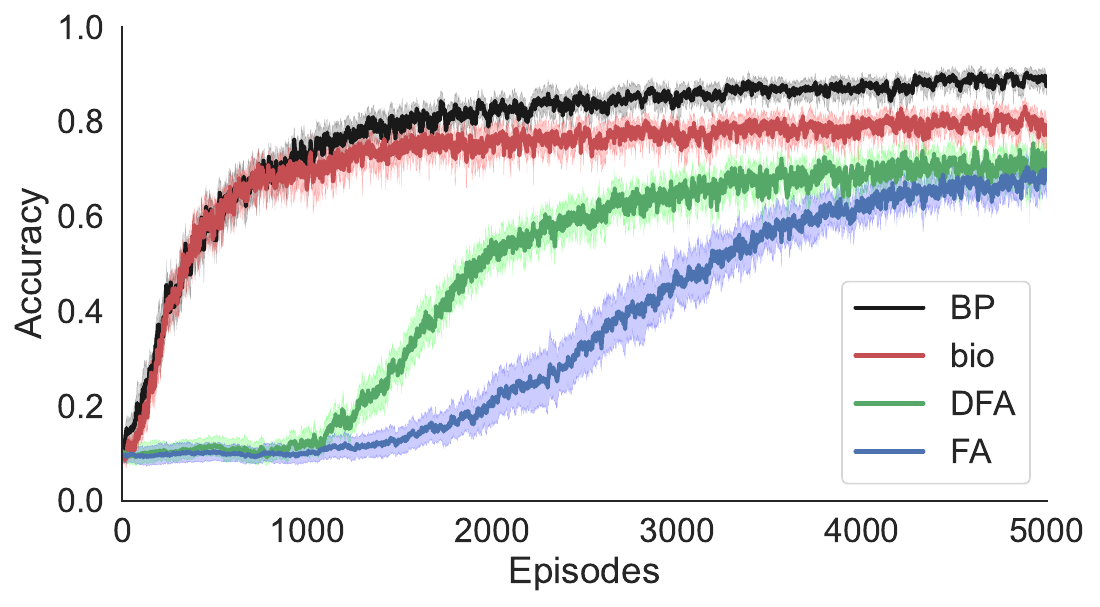}
    \caption{\textbf{Performance of benchmark learning schemes} \footnotesize \textit{while training a $5-$layer fully-connected classifier network on MNIST digits~\cite{lecun1998gradient} \nst{for a 10-way classification task. The plot demonstrates} accuracy versus the number of training data for Feedback Alignment (FA)~\cite{lillicrap2016random}, Direct Feedback Alignment (DFA)~\cite{nokland2016direct}, and backprop (BP)~\cite{rumelhart1986learning} methods, compared to the discovered biologically \revb{plausible} plasticity rule \nst{$\mathcal{F}^{bio}$ (Eq.~\ref{eq:F_bio}).}} \label{fig:DFA}}
\end{figure}

\revb{

Figure~\ref{fig:benchmarks} illustrates that the Feedback Alignment model~\cite{lillicrap2016random} is less effective than the backprop model when training deep networks with a continuous data stream. To be more precise, the backprop model begins learning immediately at the start of training, while the Feedback Alignment model takes around 2000 training data points before it starts to learn. Additionally, the rate of learning for the Feedback Alignment model is slower.

In an attempt to improve the Feedback Alignment model's performance, the Direct Feedback Alignment (DFA) method~\cite{nokland2016direct} proposed altering the backward connections to directly transmit errors from the output layer $\boldsymbol{y}_L$ to the upstream layers $\boldsymbol{y}_\ell$. The modulating signals in this modified model are calculated as 
\begin{equation*}
    \boldsymbol{e}_{\ell}= \boldsymbol{B}_{L,\ell} \boldsymbol{e}_{L}\odot\sigma'(\boldsymbol{z}_{\ell}),
    \label{eq:e_l_DFA}
\end{equation*}
with
\begin{equation*}
    \boldsymbol{e}_L=\frac{\partial \mathcal{L}}{\partial \boldsymbol{z}_{L}}.
\end{equation*}
In this formulation, $\boldsymbol{B}_{L,\ell}\in\mathbb{R}^{dim(\boldsymbol{y}_{\ell})\times dim(\boldsymbol{y}_L)}$, where $dim(\boldsymbol{y}_{\ell})$ represents the dimensionality of the activation $\boldsymbol{y}_{\ell}$.

As shown in Fig.~\ref{fig:DFA}, incorporating direct feedback connections to the Feedback Alignment method speeds up learning, and the model's accuracy improves after 1000 training data points. However, even with this modification, the network's performance is still lower than that of the backprop model. Figure~\ref{fig:DFA} further compares the DFA model with the Feedback Alignment model trained with the $\mathcal{F}^{bio}$ plasticity rule (Eq.~\ref{eq:F_bio}) and shows that the improved plasticity rule outperforms the DFA model.
}

\subsection*{Performance of the $\mathcal{F}^{bio}$}

\begin{figure}[H]
    \centering
    \includegraphics[width=.97\textwidth]{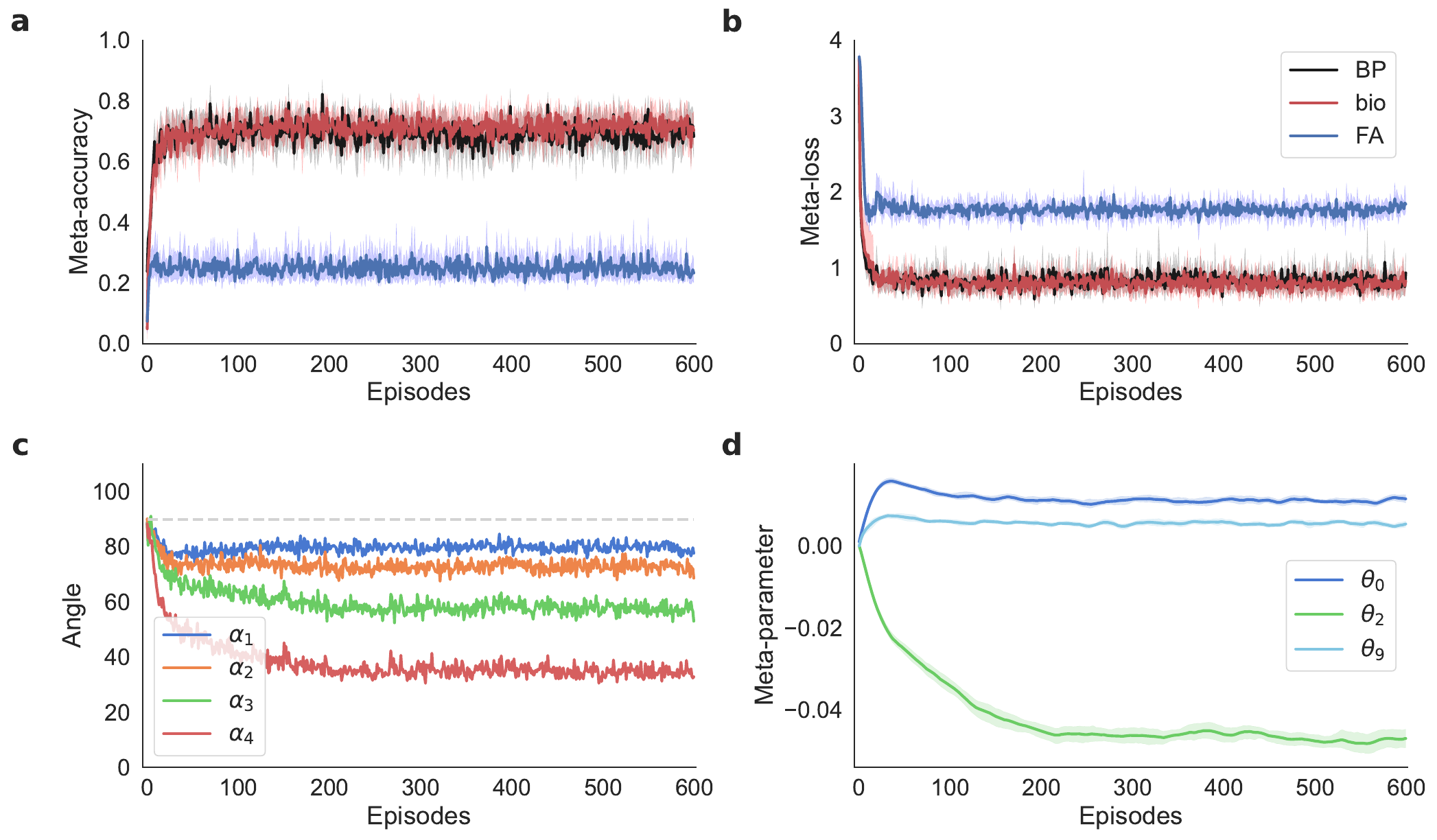}
    \caption{\textbf{Performance of the classifier network trained with $\mathcal{F}^{bio}$ plasticity rule:} \footnotesize\textit{Comparison between (a) meta-accuracy and (b) meta-loss of $\mathcal{F}^{bio}$ rule with $\mathcal{F}^{0}$ via feedback alignment (FA) and backprop (BP), (c) alignment angles between modulating signals of $\mathcal{F}^{bio}$ and backprop, and (d) convergence of the plasticity meta-parameters. While the term $\mathcal{F}^{bio}$ was discovered by regularizing the meta-loss with the penalty term in Eq.~\ref{eq:meta_L} (See Methods), $\lambda$ is set to zero for the illustrations in this figure for the uncovered rule.}
    \label{fig:F_bio}}
\end{figure}

Fig.~\ref{fig:F_bio} demonstrates the classifier's performance with $\mathcal{F}^{bio}$ within \revb{$600$} iterations of the meta-optimizer. Comparing the loss and accuracy of $\mathcal{F}^{bio}$ with $\mathcal{F}^{0}$ via feedback alignment in Fig.~\ref{fig:F_bio}a and Fig.~\ref{fig:F_bio}b, respectively, shows a significant boost in learning through $\mathcal{F}^{bio}$. Figure~\ref{fig:F_bio}c further shows improvement in the alignment of the modulating signals with those of the backprop. These angles are reduced the most in the deeper layers. Lastly, Fig.~\ref{fig:F_bio}d illustrates the progress of the meta-parameters. We observe that the plasticity coefficients converge in about 200 episodes.

\subsection*{Data flow in $\mathcal{F}^{eHebb}$\label{app:F_eHebb}}

\begin{figure}[H]
    \centering
    \includegraphics[width=.91\textwidth]{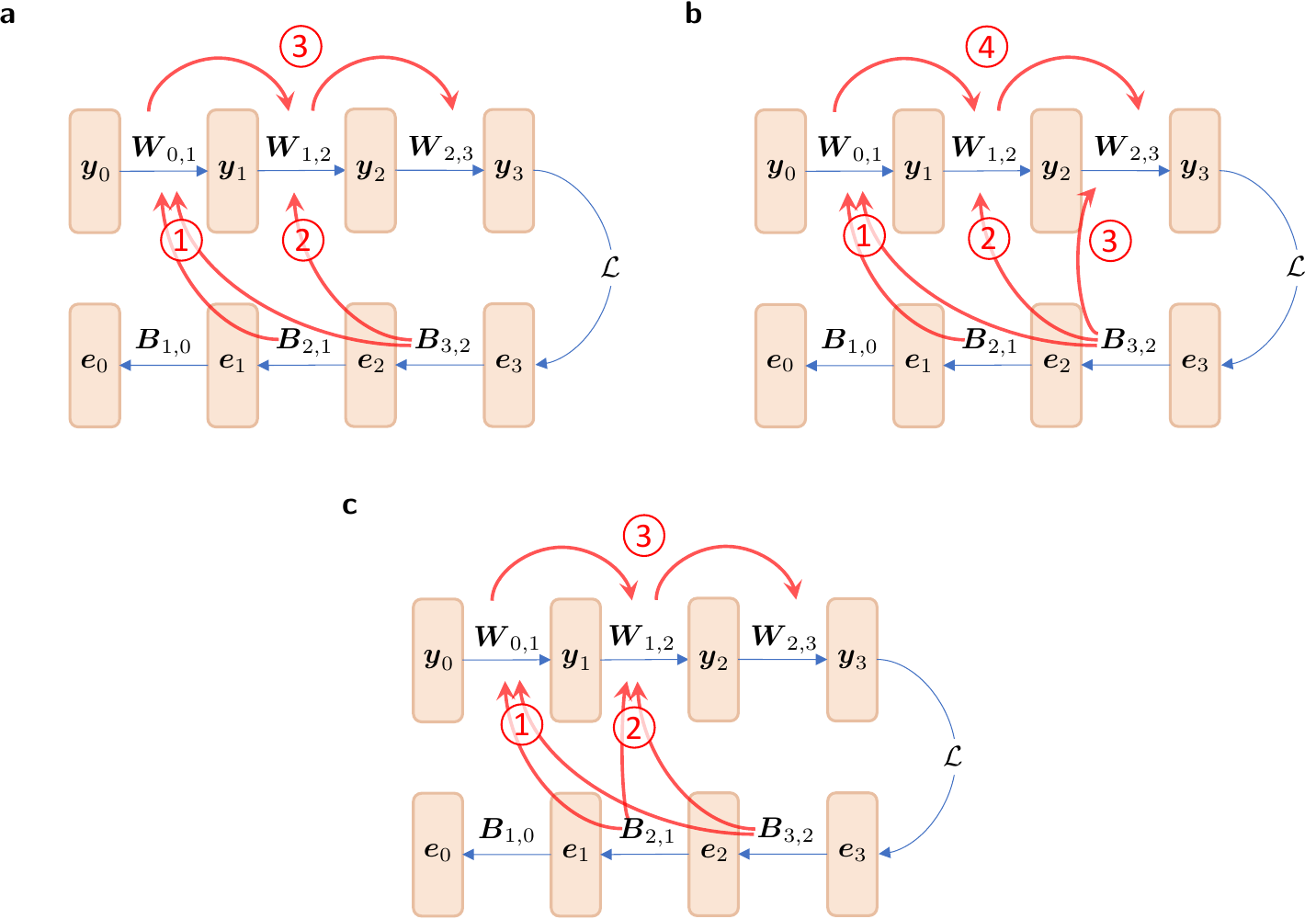}
    \caption{\textbf{Interactions between feedback and forward pathways using $\mathcal{F}^{eHebb}$:} \footnotesize \textit{(a) All layers trained with \reva{the rule $\mathcal{F}(\boldsymbol{\Theta})=\theta_0\mathcal{F}^{0}$} via feedback alignment. \reva{Information from $\boldsymbol{B}_{3,2}$ and $\boldsymbol{B}_{2,1}$ is transmitted to $\boldsymbol{W}_{0,1}$ through the $\mathcal{F}^{0}$ plasticity rule (\emph{\textcircled{\scriptsize 1}}), which then passes on to $\boldsymbol{W}_{1,2}$ and $\boldsymbol{W}_{2,3}$ (\emph{\textcircled{\scriptsize 3}}). Meanwhile, information from $\boldsymbol{B}_{3,2}$ is transmitted to $\boldsymbol{W}_{1,2}$ (\emph{\textcircled{\scriptsize 2}}), which is then propagated to $\boldsymbol{W}_{2,3}$ after the forward propagation.} (b) $\boldsymbol{W}_{2,3}$ is updated using \reva{$\mathcal{F}^{eHebb}(\boldsymbol{\Theta})=\theta_0\mathcal{F}^{0}+\theta_2\mathcal{F}^{2}$}, while $\boldsymbol{W}_{0,1}$ and $\boldsymbol{W}_{1,2}$ are trained with \reva{the rule $\mathcal{F}(\boldsymbol{\Theta})=\theta_0\mathcal{F}^{0}$} via feedback alignment. \reva{Plasticity rule $\mathcal{F}^{0}$ transmits information from $\boldsymbol{B}_{2,1}$ and $\boldsymbol{B}_{3,2}$ to $\boldsymbol{W}_{0,1}$ (\emph{\textcircled{\scriptsize 1}}) and from $\boldsymbol{B}_{3,2}$ to $\boldsymbol{W}_{1,2}$ (\emph{\textcircled{\scriptsize 2}}). This information is propagated to their downstream layers after the forward path (\emph{\textcircled{\scriptsize 4}}). Concurrently, an additional channel established by $\mathcal{F}^{2}$ explicitly propagates the information from $\boldsymbol{B}_{3,2}$ to $\boldsymbol{W}_{2,3}$ (\emph{\textcircled{\scriptsize 3}}).} (c) $\boldsymbol{W}_{0,1}$ and $\boldsymbol{W}_{2,3}$ use \reva{the plasticity rule $\mathcal{F}(\boldsymbol{\Theta})=\theta_0\mathcal{F}^{0}$} via feedback alignment, and $\boldsymbol{W}_{1,2}$ utilizes \reva{$\mathcal{F}^{eHebb}(\boldsymbol{\Theta})=\theta_0\mathcal{F}^{0}+\theta_2\mathcal{F}^{2}$}. \reva{$\mathcal{F}^{0}$ communicates information from $\boldsymbol{B}_{2,1}$ and $\boldsymbol{B}_{3,2}$ to $\boldsymbol{W}_{0,1}$ (\emph{\textcircled{\scriptsize 1}}), which then is propagated to the downstream layers (\emph{\textcircled{\scriptsize 3}}). Meanwhile, the $\mathcal{F}^{0}$ rule in $\mathcal{F}^{eHebb}$ disseminates information from $\boldsymbol{B}_{3,2}$ to $\boldsymbol{W}_{1,2}$, while $\mathcal{F}^{2}$ in $\mathcal{F}^{eHebb}$ establishes a direct route to transmit information from $\boldsymbol{B}_{2,1}$ to $\boldsymbol{W}_{1,2}$ (\emph{\textcircled{\scriptsize 2}}). The ensuing forward propagation from $\boldsymbol{W}_{1,2}$ to the downstream layers continues as usual.} In all graphs, blue arrows represent the propagation of data through the forward or backward path, while the red arrow represents the flow of information from the backward pathway to the forward connections.} \label{fig:F_eHebb_flow_3L}}
\end{figure}

Table~\ref{tab:F_eHebb_alignment} demonstrates the effect of the Hebbian-like error plasticity term (Eq.~\ref{eq:F_eHebb}) on the alignment angles of the modulator signals. Here, we explain these improvements by illustrating $\mathcal{F}^{eHebb}$'s influence on the feedback pathway's interactions with the forward path. To set the baseline, Fig.~\ref{fig:F_eHebb_flow_3L}a employs the plasticity rule
\begin{equation}
    \mathcal{F}^{0}\left(\boldsymbol{\Theta}\right)=-\theta_0 \boldsymbol{e}_{\ell} \boldsymbol{y}_{\ell-1}^{T}
    \label{eq:F_pseudo}
\end{equation}
to train the network (row 1 in Tab.~\ref{tab:F_eHebb_alignment}), where $\boldsymbol{e}_{\ell}$ is transmitted through random feedback pathways. First, information from backward connections $\boldsymbol{B}_{2,1}$ and $\boldsymbol{B}_{3,2}$ (through $\boldsymbol{B}_{2,1}$) flows into $\boldsymbol{W}_{0,1}$ via Eqs.~\ref{eq:e_l_backprop} and~\ref{eq:F_pseudo}. Similarly, information from $\boldsymbol{B}_{3,2}$ flows into $\boldsymbol{W}_{1,2}$ during the weight update. Then in the forward pass, information from $\boldsymbol{W}_{0,1}$ and $\boldsymbol{W}_{1,2}$ are propagated forward into $\boldsymbol{W}_{2,3}$. Table~\ref{tab:F_eHebb_alignment} shows that this flow of information does not sufficiently adjust $\boldsymbol{W}$ for a good alignment of the teaching signals, particularly in online training with limited data.

In Fig.~\ref{fig:F_eHebb_flow_3L}b, we add the Hebbian-style error term to update $\boldsymbol{W}_{2,3}$ using $\mathcal{F}^{eHebb}$, while the rest of the network is trained with $\mathcal{F}^{0}$ through feedback alignment (Tab.~\ref{tab:F_eHebb_alignment}, row 3). The information flow to $\boldsymbol{W}_{0,1}$ and $\boldsymbol{W}_{1,2}$ stays the same; however, $\mathcal{F}^{eHebb}$ introduces an auxiliary information channel from $\boldsymbol{B}_{3,2}$ to $\boldsymbol{W}_{2,3}$. As presented in Tab.~\ref{tab:F_eHebb_alignment}, this supplementary channel results in a better alignment of $\boldsymbol{e}_2$ with the corresponding error vector transmitted via backprop.

Figure~\ref{fig:F_eHebb_flow_3L}c repeats this experiment with $\boldsymbol{W}_{1,2}$ updated using $\mathcal{F}^{eHebb}$ while other layers are updated with $\mathcal{F}^{0}$ with feedback alignment (Tab.~\ref{tab:F_eHebb_alignment}, row 2). Although $\boldsymbol{W}_{0,1}$ is updated with the same flow of information as Fig.~\ref{fig:F_eHebb_flow_3L}b, there is a new flow from $\boldsymbol{B}_{2,1}$ to $\boldsymbol{W}_{1,2}$, which improves $\boldsymbol{e}_1$'s alignment. Note that better alignment of $\boldsymbol{e}_1$ results in more backprop-like weight update, which subsequently improves data propagation to the downstream layers. As a result, the alignments in the downstream layers are slightly improved as well, even with the vanilla $\mathcal{F}^{0}$ plasticity rule with feedback alignment updating them. This behavior is similar to the reduced alignment angles in Fig.~\ref{fig:F_Oja}c, where $\mathcal{F}^{Oja}$ positively affects the alignments by improving the forward data propagation.

\subsection*{Expectation of Hebbian-style error-based plasticity \label{app:F_eHebb_math}}

Assume that the entries of $\boldsymbol{B}_{\ell+1,\ell}$ are i.i.d. with expectation zero  and independent  from the entries of $\boldsymbol{e}_{\ell+1}$. Also assume that the entries of $\boldsymbol{e}_\ell$ have variance $\sigma_\ell^2$.
In this Supplementary section, we show that
\begin{equation}\label{E:AppResult}
\mathbb E\left[\left.\boldsymbol{e}_\ell \boldsymbol{e}_{\ell-1}^T\, \right|\,\boldsymbol{B}_{\ell,\ell-1}\right]=\sigma_\ell^2 \boldsymbol{B}_{\ell,\ell-1}^T
\end{equation}
We must first show that  $\mathbb{E}[(\boldsymbol{e}_\ell)_j)]=0$ and $\mathbb{E}[(\boldsymbol{e}_\ell)_i(\boldsymbol{e}_\ell)_j)]=0$ when $i\ne j$ by computing
\[
\begin{aligned}
\mathbb{E}\left[\left(\boldsymbol{e}_\ell\right)_j\right]&=\mathbb{E}\left[\left(\sum_{k} \left(\boldsymbol{B}_{\ell+1,\ell}\right)_{j,k}\left(\boldsymbol{e}_{\ell+1}\right)_{k}\right)\right]\\
&=\sum_k \mathbb{E}\left[\left(\boldsymbol{B}_{\ell+1,\ell}\right)_{j,k}\right]\mathbb{E}\left[\left(\boldsymbol{e}_{\ell+1}\right)_{k}\right]\\
&=0.
\end{aligned}
\]
Now, assume that $i\ne j$ and compute
\[
\begin{aligned}
\mathbb{E}\left[(\boldsymbol{e}_\ell)_i(\boldsymbol{e}_\ell)_j\right]&=\mathbb{E}\left[\left(\sum_k (\boldsymbol{B}_{\ell+1,\ell})_{i,k}\right)(\boldsymbol{e}_{\ell+1})_k\left(\sum_{k'} (\boldsymbol{B}_{\ell+1,\ell})_{j,k'}\right)(\boldsymbol{e}_{\ell+1})_{k'}\right]\\
&\revb{=\sum_{k,k'}\left(\mathbb{E}\big[\left(\boldsymbol{B}_{\ell+1,\ell}\right)_{i,k}\left(\boldsymbol{B}_{\ell+1,\ell}\right)_{j,k'}\big]\mathbb{E}\big[\left(\boldsymbol{e}_{\ell+1}\right)_k\left(\boldsymbol{e}_{\ell+1}\right)_{k'}\big]\right)}\\
&=0
\end{aligned}
\]
where the last line follows from the assumptions that $i\ne j$ and $\boldsymbol{B}_{\ell+1,\ell}$ has independent entries. 

Now we can derive Eq.~\eqref{E:AppResult} as follows
\[
\begin{aligned}
\mathbb E\left[\left.\left(\boldsymbol{e}_\ell \boldsymbol{e}^T_{\ell-1}\right)_{i,k}\, \right|\,\boldsymbol{B}_{\ell,\ell-1}\right]&= \mathbb E\left[\left.\left(\boldsymbol{e}_\ell \boldsymbol{e}_{\ell}^T\boldsymbol{B}^T_{\ell,\ell-1}\right)_{i,k}\,\right|\,\boldsymbol{B}_{\ell,\ell-1}\right]\\
&=\mathbb E\left[\left.\sum_{j=1}^n \left(\boldsymbol{e}_\ell \boldsymbol{e}^T_{\ell}\right)_{i,j}\left(\boldsymbol{B}^T_{\ell,\ell-1}\right)_{j,k}\,\right|\,\boldsymbol{B}_{\ell,\ell-1}\right]\\
&=\sum_{j=1}^n\mathbb E\left[\left. \left(\boldsymbol{e}_\ell\right)_i\left( \boldsymbol{e}_{\ell}\right)_{j}\left(\boldsymbol{B}^T_{\ell,\ell-1}\right)_{j,k}\,\right|\,\boldsymbol{B}_{\ell,\ell-1}\right]\\
&=\mathbb E\left[\left. \left(\boldsymbol{e}_\ell\right)_i\left( \boldsymbol{e}_{\ell}\right)_{i}\left(\boldsymbol{B}^T_{\ell,\ell-1}\right)_{i,k}\,\right|\,\boldsymbol{B}_{\ell,\ell-1}\right]\\
&=\sigma_\ell^2 \left(\boldsymbol{B}^T_{\ell,\ell-1}\right)_{i,k}
\end{aligned}
\]
The last two lines follow from the fact that whenever $i\ne j$, the expectation is equal to zero. Eq.~\eqref{E:AppResult} follows directly.

\begin{figure}[H]
    \centering
    \includegraphics[width=.475\textwidth]{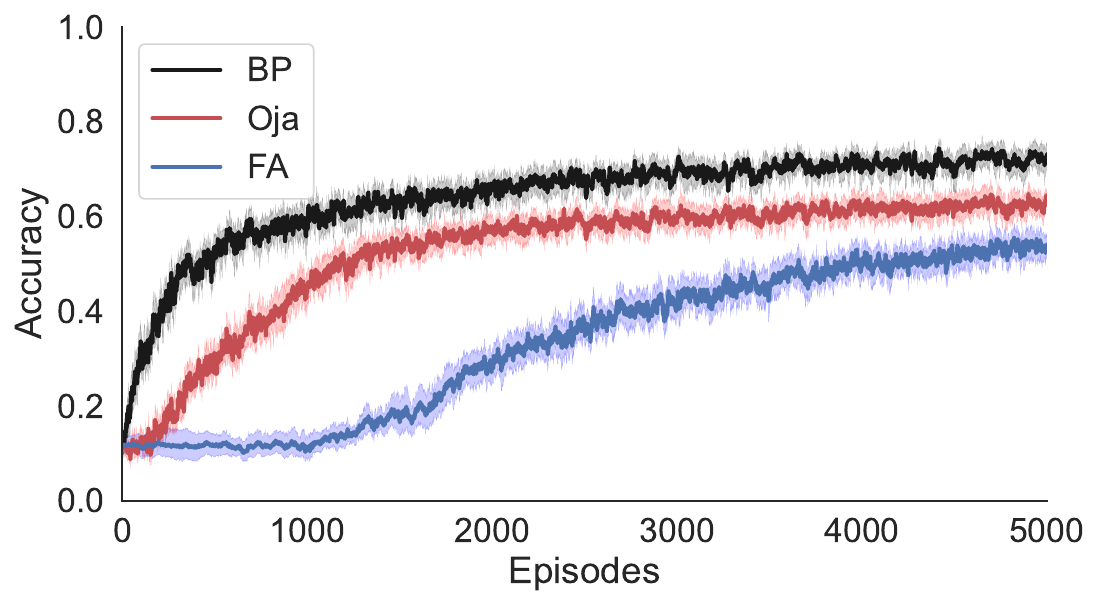}
    \caption{\reva{\textbf{Performance of benchmark learning schemes} \footnotesize \textit{while training a $5-$layer fully-connected classifier network on FashionMNIST dataset~\cite{xiao2017fashion} for a 10-way classification task. The plot demonstrates accuracy versus the number of training data for Feedback Alignment (FA)~\cite{lillicrap2016random} and backprop (BP)~\cite{rumelhart1986learning} methods, compared to $\mathcal{F}^{Oja}$ (Eq.~\ref{eq:F_Oja}).}}} \label{fig:FashionMNIST}
\end{figure}

\reva{
\subsection*{Performance of the $\mathcal{F}^{Oja}$ on FashionMNIST}

Section 2.2.2 examines how Oja's rule improves learning in the Feedback Alignment model (Fig.~\ref{fig:F_Oja}). In this section, we demonstrate the effectiveness of Oja's rule on a different dataset by using the FashionMNIST~\cite{xiao2017fashion} to train a classifier model. Figure~\ref{fig:FashionMNIST} illustrates that introducing the Oja's rule (Eq.~\ref{eq:F_Oja}) substantially enhances learning across different datasets when the model is trained with random feedback connections.
}

\begin{figure}[H]
    \centering
    \includegraphics[width=.97\textwidth]{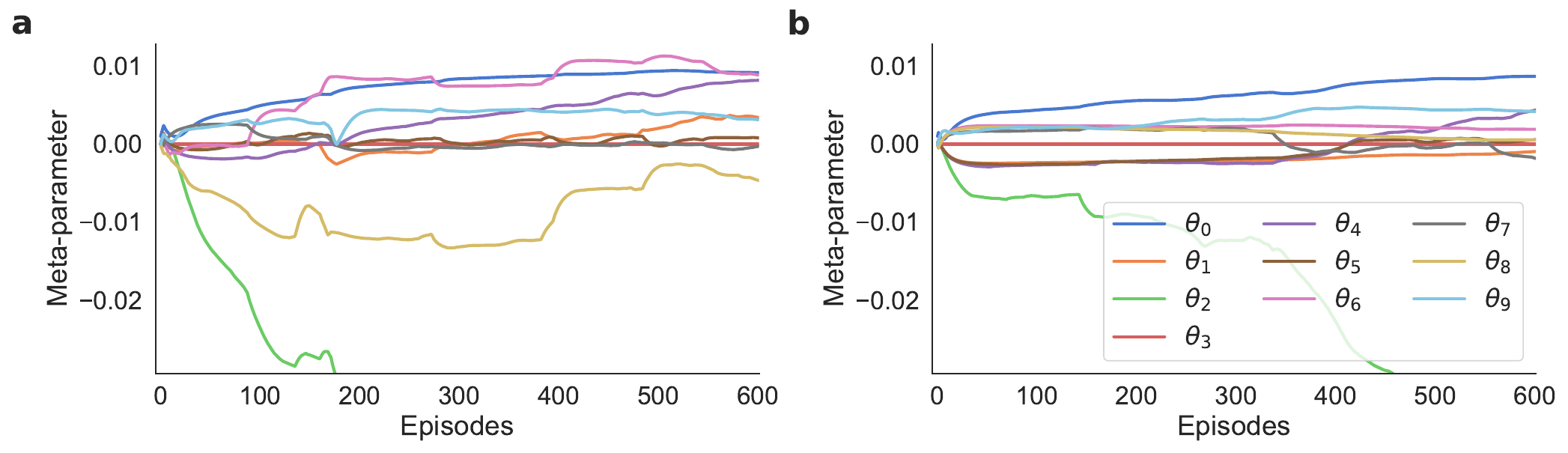}
    \caption{\revb{\textbf{L1 improves feature selection in the meta-learning model:} \footnotesize \textit{Performance of different penalization methods while training a $5-$layer fully-connected classifier network on EMNIST digits~\cite{cohen2017emnist} with online learning. Evolution of meta-parameters for the pool of learning rules defined in section~\ref{app:F_bio} using (a) no penalization, (b) L2 penalized meta-loss (Eq.~\ref{eq:meta_L2}). 
    } \label{fig:penalization}}}
\end{figure}

\revb{
\subsection*{Performance of alternative penalization methods}

In Sec.~\ref{sec:meta_plasticity}, we proposed using L1 regularization on the meta-loss to decrease redundancy within the update rules. As shown in Fig.~\ref{fig:F_pool}d, this technique leads to a sparser set of meta-parameters and acts as a model selection method, identifying the most effective plasticity rules. 

In Fig.~\ref{fig:penalization}, we examine the impact of alternative regularization methods on the meta-learning algorithm by comparing the performance of models with no regularization and L2 regularization. When using no regularization in the meta-learning, the algorithm eliminates update terms negatively impacting the learning. However, another set of plasticity rules may individually improve the results, but when these rules are considered in a set, other terms may be more beneficial for the optimization process. Nevertheless, the model still includes them in the final meta-optimized learning rule. As seen in Fig.~\ref{fig:penalization}a, the model has identified seven plasticity terms, making it impractical to investigate each of these terms individually.

As an alternative, Fig.~\ref{fig:penalization}b shows the results of using L2 regularization
\begin{equation}
    \mathcal{L}_{meta}(\boldsymbol{\Theta})=\mathcal{L}(f_{\boldsymbol{W}}(\boldsymbol{X}_{query}), \boldsymbol{Y}_{query}) + \lambda \|\boldsymbol{\Theta}\|_2.
    \label{eq:meta_L2}
\end{equation}
Unlike L1 regularization, L2 tends to decrease all parameters but does not return sparse solutions and is unsuitable for feature selection. In other words, even though L2 regularization reduces the values of all parameters, it does not eliminate the redundant or less influential plasticity terms with large meta-parameters from the final solution.

}

\begin{figure}[H]
    \centering{
    \includegraphics[width=.475\textwidth]{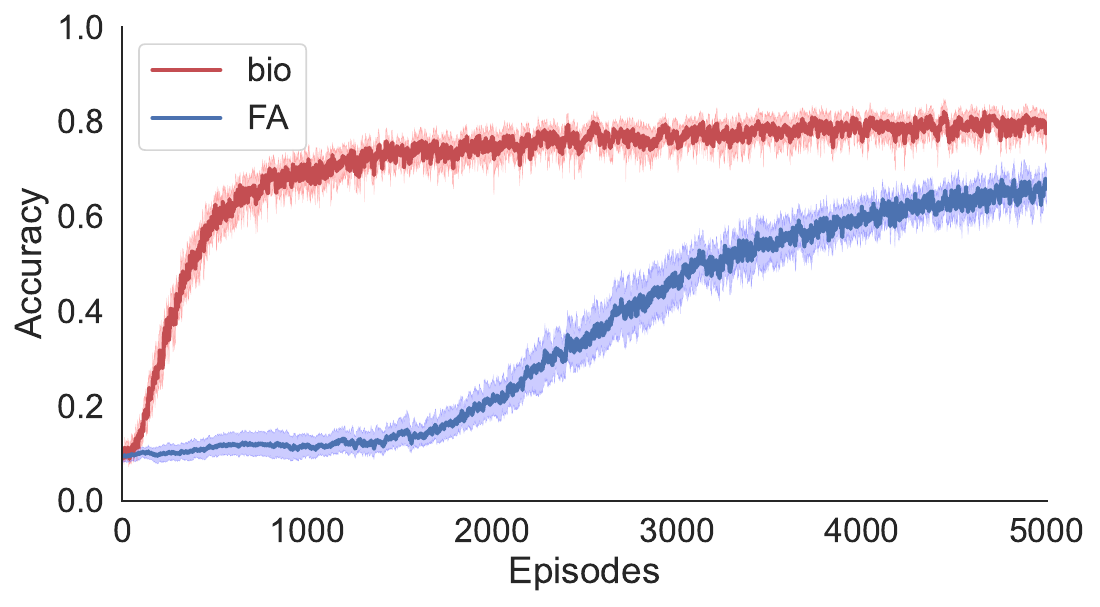}
    }
    \caption{\revb{\textbf{$\mathcal{F}^{bio}$ trains effectively under different initialization of the feedback:} \textit{Accuracy of a 5-layer classifier network trained on MNIST dataset~\cite{lecun1998gradient} to perform a 10-way classification task using Feedback Alignment (FA)~\cite{lillicrap2016random} compared to the proposed $\mathcal{F}^{bio}$ plasticity rule (bio) outlined in Eq.~\ref{eq:F_bio}. The backward connections were initialized in both tests using the normal Xavier initialization method (Eq.~\ref{eq:Xav_norm}).}}
    }\label{fig:fbk_init}
\end{figure}

\revb{
\subsection*{Performance of alternative backward initialization}

As mentioned in Sec.~\ref{app:models}, the Xavier initialization method was used to randomly sample forward and backward connections from a uniform distribution 
\begin{equation}
    \boldsymbol{B}_{\ell+1, \ell}, \boldsymbol{W}_{\ell, \ell+1}\sim \mathcal{U}\left(-\sqrt{\frac{6}{dim(\boldsymbol{y}_{\ell})+dim(\boldsymbol{y}_{\ell+1})}}, \sqrt{\frac{6}{dim(\boldsymbol{y}_{\ell})+dim(\boldsymbol{y}_{\ell+1})}}\right)
    \label{eq:Xav_uni}
\end{equation}
throughout the study, where $dim(\boldsymbol{y}_{\ell})$ is the dimension of the activation $\boldsymbol{y}_{\ell}$. Nevertheless, the findings presented in this work do not depend on the initialization method of the backward connections.


To illustrate this, we conducted an experiment where we employed the normal Xavier initialization method
\begin{equation}
    \boldsymbol{B}_{\ell+1, \ell}\sim \mathcal{N}\left(0, \frac{2}{dim(\boldsymbol{y}_{\ell})+dim(\boldsymbol{y}_{\ell+1})}\right)
    \label{eq:Xav_norm}
\end{equation}
to sample initial values for the backward connections. The forward connections were initialized using a uniform distribution as before (Eq.~\ref{eq:Xav_uni}). Figure~\ref{fig:fbk_init} shows that the proposed $\mathcal{F}^{bio}$ plasticity rule can successfully train the model using different methods for initializing the backward connections.
}

\begin{figure}[H]
    \centering
    \includegraphics[width=.97\textwidth]{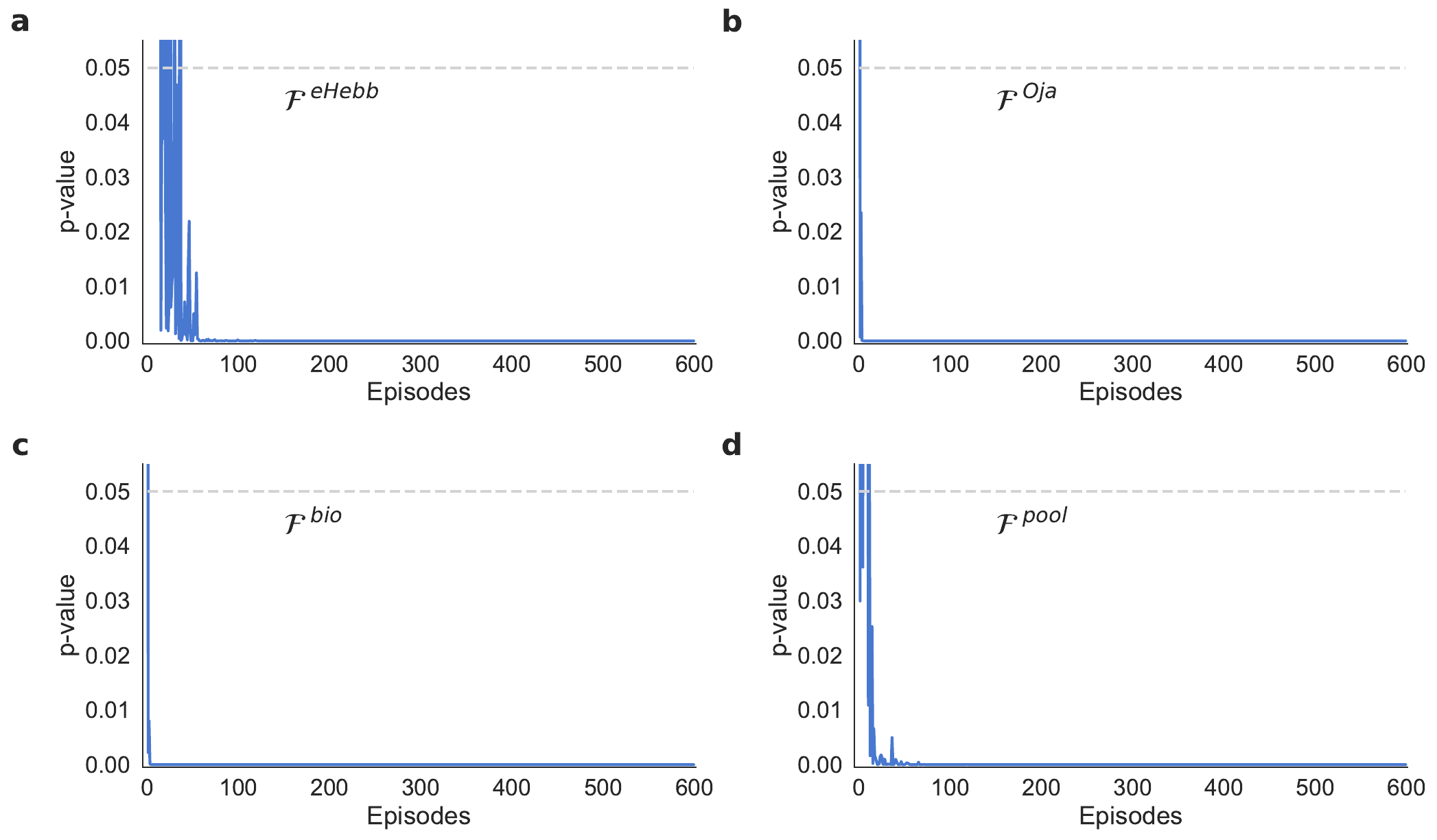}
    \caption{\revc{\textbf{The performance gain obtained with the modified plasticity rules is statistically significant:} \textit{The p-value of the one-sided Mann-Whitney test over 600 meta-optimization episodes, comparing samples from trials using the FA method to those using (a) $\mathcal{F}^{eHebb}$, (b) $\mathcal{F}^{Oja}$, (c) $\mathcal{F}^{bio}$, and (d) $\mathcal{F}^{pool}$ plasticity rules.}}
    \label{fig:MannWhitney}}
\end{figure}

\revc{

\subsection*{Inter-treatment variation}

Throughout the paper, we examine the variations within each plasticity rule by calculating the confidence intervals. To determine if the improvements in accuracy are statistically significant, we use the Mann-Whitney U test to compare two sets of data: the accuracy of trials using the FA method and the modified plasticity rule. Samples are taken at the end of each episode and represent the accuracy of the model trained with different initial weights and feedback connection values. We chose the Mann-Whitney U test over the t-test as it does not assume a Gaussian distribution within the groups.

We begin by hypothesizing that the FA method trial samples show lower accuracy than that of the modified plasticity rule. We utilize 20 samples from each group. The results, illustrated in Fig.~\ref{fig:MannWhitney}, indicate that the p-value falls below $5\%$ within fewer than 100 episodes in every example. Our findings indicate strong evidence against the null hypothesis, providing statistical support for the performance gain using the proposed plasticity rules.
}

\end{document}